\documentclass[prx,a4paper,aps,onecolumn,superscriptaddress,11pt]{quantumarticle}

\usepackage[utf8]{inputenc}
\usepackage[english]{babel}
\usepackage[T1]{fontenc}
\usepackage{amsmath, amssymb}  
\usepackage{hyperref}
\usepackage{pbox}
\usepackage[numbers]{natbib}

\usepackage{tikz}
\usetikzlibrary{calc, through, arrows}
\usepackage{lipsum}
\usepackage{graphicx}
\DeclareGraphicsExtensions{.png}
\usepackage{caption}
\usepackage{subcaption}

\newcommand{\ov}{\overline}

\newcommand{\mbb}{\mathbb}

\newcommand{\p}{\prime}
\newcommand{\ra}{\rangle}
\newcommand{\la}{\langle}

\newcommand{\re}{\textup{Re}}

\newcommand{\wh}{\widehat}

\newcommand{\Tr}{\textup{Tr}}

\newtheorem{theorem}{Theorem}
\newtheorem{corollary}[theorem]{Corollary}

\newtheorem{assumption}{Assumption}
\newtheorem{definition}{Definition}
\newtheorem{conjecture}[theorem]{Conjecture}

\newtheorem{remark}{Remark}

\begin{document}

\title{Quantum Cryptography with Weak Measurements}
\date{\today}
\author{James E. Troupe}
\affiliation{Center for Quantum Research, Applied Research Laboratories, University of Texas at Austin, Austin, TX 78713 USA}
\affiliation{Institute for Quantum Studies, Chapman University, Orange, CA 92866 USA}
\email{jtroupe@arlut.utexas.edu}
\author{Jacob M. Farinholt}
\email{jacob.farinholt@navy.mil}
\thanks{Both authors contributed equally to this paper.}
\affiliation{Strategic and Computing Systems Department, Naval Surface Warfare Center, Dahlgren Division, Dahlgren, VA 22448 USA}
\maketitle

\begin{abstract}
In this article we present a new prepare and measure quantum key distribution protocol that decouples the necessary quantum channel error estimation from its dependency on sifting, or otherwise post-selecting, the detection outcomes. Rather than estimating Eve's coupling to the quantum channel from the statistics of the sifted key, we infer this information from weak measurements made equally on all of the received photons immediately prior to post-selection by the photon detectors. We prove that the accuracy of the weak measurement parameter estimation is robust to reasonable device imperfections, even in an adversarial environment, and hence the asymptotic security of this protocol can be inferred from the security analysis of BB84. In addition to eliminating detector basis-dependent attacks, such as detector blinding, we demonstrate that this new prepared and measure QKD protocol is immune to a very powerful class of measurement-side device attacks that also allow an adversary control of the weak measurement outcomes given two modest requirements placed on the measurement-side devices. Finally, we compare the asymptotically achievable secure key rate of a decoy state version of the weak measurement protocol and show it is essentially equal to that of BB84 with decoy states and significantly higher than MDI-QKD for realistic system parameters. 
\end{abstract}

\section{Introduction}
In the asymptotic security analysis of QKD protocols, one prevailing assumption is that the bit error rates can be ascertained to arbitrary precision through standard post-selection measurements and classical communication. For example, in \cite{GLLP}, the accuracy of the bit error rate was taken for granted, whereas bounds were calculated for the variance of the phase error rate from the bit error rate due to device imperfections. 

To estimate the error rates of the quantum channel for prepare and measure QKD protocols, Alice and Bob sift the detection outcomes by throwing out the cases where Bob's measurement basis is different from Alice's preparation basis. A randomly chosen subset of the sifted key is then used to estimate the channel error rate. An implicit assumption of this procedure is that the signals that contributed to the sifted key are statistically representative of all of the signals transmitted through the quantum channel. However, there exist strongly basis dependent attacks that can exploit the behavior of the detectors to invalidate this assumption of fair sampling from the quantum channel, e.g detector blinding attacks. Detector blinding attacks have been successfully demonstrated on both prepare-and-measure and entanglement based QKD protocols utilizing detector behavior \cite{Lyd10,Gerh11,Sajeed16,Makarov16}. Another type of detector based attack is one that exploits individual characteristics of the detectors, e.g. timing response \cite{Lamas07}, that can cause side-channels that leak information about the detection outcomes. This type of side-channel is important to characterize and limit in any QKD system, however they are generally regarded as less dangerous than detector attacks that use active control of the detectors to target the detection basis. This is because, for almost all protocols, the latter type of attack can completely eliminate the ability to detect eavesdropping. 

In an effort to eliminate all detector side-channel attacks, so-called Measurement-Device-Independent (MDI) protocols \cite{LoCurQi12,Bing13,Curty14,Tang14,Tittel15} were developed that replace Bob's detectors with an additional source, and insert a Bell measurement half-way between Alice and Bob. In so doing, these protocols replace the dependence of the security on Alice's source and Bob's detectors with its dependence on both Alice and Bob's sources. In addition to MDI-QKD, a recent proposal known as Detector-Device-Independent QKD (DDI-QKD) attempted to close the detector side-channel weakness of prepare and measure protocols by employing a randomized linear optical network between the quantum channel and the receiver's detectors \cite{Gonzalez15}. Unfortunately, DDI-QKD has very recently been shown to in fact be vulnerable to detector blinding attacks \cite{Sajeed16}, leaving MDI-QKD as currently the most feasible option for eliminating all detector side-channel attacks, with detector blinding arguably being the most serious of these. 

In this article we introduce a new method to utilize single qubit preparation and measurement to perform secure quantum key distribution that explicitly removes the dependency of channel estimation on the detected bit values. We propose to estimate the phase and bit errors of the channel by the use of weak measurements of particular observables performed immediately preceding Bob's final post-selections. As we will show, this approach decouples the error analysis from Bob's detection outcomes, and therefore removes the assumption of fair sampling implicit in all other prepare and measure QKD protocols by using the weak measurement results from \emph{all} received signals equally to perform security parameter estimation. While we cannot call this protocol measurement-device-independent, we will nevertheless demonstrate that as long as we can place reasonable bounds on the leakage of information from the weak measurement device, and we can enforce that Eve is unable to manipulate the signal between the weak measurement and the final detection, then it is possible to obtain a secure key \emph{even if Eve completely controls the weak measurement results} under a very strong class of measurement-device based attacks. Since the difference between the weak measurement protocol and conventional single qubit QKD protocols is the method of error estimation, the weak measurement protocol does not remove the potential for side-channels that leak sifted key via differences in detector timing. Additionally, as is true for even fully ``Device-Independent'' QKD protocols \cite{Barrett05}, the new weak measurement protocol does not prevent memory attacks that use compromised components to record and leak generated key \cite{Barrett13}. 


The rest of the paper is arranged as follows. Section \ref{WM-Background} will provide the necessary background material on weak measurements. Section \ref{Sec:WM-Protocol} will lay out the general framework of a QKD protocol using weak measurements. Section \ref{Sec:PE} sketches an entanglement distillation version of the protocol in order to rigorously connect the security of the protocol to the accuracy of the error parameter estimation. This is followed by a method to ascertain the error parameters from the expectation values determined by the weak measurement results. Section \ref{Sec:WM-disturbance} demonstrates the negligible effect these weak measurements have on the key rate. In Section \ref{Sec:WM-Security} we perform a full (asymptotic) security analysis of the protocol in the presence of several device imperfections and calculate the effect of these imperfections on key rates. In Section \ref{Sec:MDI-security}, we consider the most adversarial case in which the weak measurements take place inside a black box in the possession of Eve and operated internally by Fred, an agent of Eve. We assume that Fred has limited ability to communicate with Eve; however, Eve has unlimited ability to communicate with Fred. We conjecture an optimal attack strategy for Eve in this scenario, and verify the protocol is secure against this attack even when Eve's interaction with the quantum channel is strongly basis dependent and Eve has complete control of the weak measurement results themselves. In Section \ref{Sec:Implementation} we briefly discuss possible implementations of the required weak measurements. And finally, in Section \ref{Sec:Performance} we compare the asymptotically achievable secure key rates of the new weak measurement protocol to that of BB84 \cite{BB84,LMC2005} and MDI-QKD \cite{LoCurQi12} for realistic system parameters. 

\section{\label{WM-Background}Background on Weak Measurements}

The concept of weak measurements has a long and interesting history with its origins in the investigation of the foundations of quantum measurement.  There are two main approaches to weak measurements roughly divided by their intended use: continuous and discrete (or quasi-instantaneous) weak measurements.  Continuous weak measurements developed from the theory of open quantum systems \cite{CBR91, Car93, BS92, Wiseman02}, and find great utility today in methods for feedback and adaptive control of quantum systems, particularly for quantum information processing \cite{WCDJMS14, CCetal17}. What we will refer to as ``discrete'' weak measurements were introduced by Aharonov and colleagues in 1988 \cite{AAV1988, AV90} as a method of providing experimental access to weak values, the values of observables of a quantum system at an intermediate time between pre-selection and post-selection by initial and final quantum states. While we will not be explicitly using any of the properties of weak values themselves in this article, the QKD protocol presented was inspired by them. Accordingly, throughout this article we will be utilizing the discrete interaction version of weak measurements. 

The basic idea of a weak measurement is for the measuring system, which we refer to as the measurement device, to itself be treated as quantum mechanical, with a quantum state that is a well-defined superposition of initial positions (and accordingly relatively certain momentum), and then engineer a weak interaction between the measurement device and the target quantum system over a brief time interval.  The result is that the measurement device’s position wave function is evolved by a small translation proportional to the weak value of the pre- and post-selected target system. In addition, since the interaction is weak, the measuring device and target system are only very weakly entangled so that upon a projective measurement of the measurement device’s position, the target system is only very rarely projected into an orthogonal state. Unlike the standard strong measurement of a quantum mechanical observable which disturbs the measured system and ``collapses'' its state into an eigenstate of the observable, a weak measurement does not appreciably disturb the quantum system. This is possible because very little information about the observable is extracted in a single weak measurement. 

It is not at all necessary that the target system be post-selected when performing weak measurements.  In that case, instead of registering a translation proportional to the real component of the weak value, the average translation will be an amount proportional to the expectation value of the observable given the target system’s initial state.  This means that we can use weak measurements of a large set of identically prepared quantum systems to estimate the expectation value of an observable without significantly disturbing the individual systems' quantum states.  

More formally, weak measurements can be described using the von Neumann model of quantum
measurement \cite{vonNeumann55}. Consider an observable $\widehat{A}$ pertaining to a quantum system pre-selected to be in the state $\left\vert \psi_{in}\right\rangle $. The Hamiltonian describing the
interaction between the system and measuring device (MD) is
\begin{equation}
\widehat{H}_{int}=g(t)\widehat{P}_{MD} \otimes \widehat{A},
\end{equation}
where $\widehat{P}_{MD}$ is the momentum operator for the MD. The evolution operator for the system and MD\ is then given by 
\begin{equation}
\widehat{U}_{wm} = \exp\left(-i\int_{t_{0}-\epsilon}^{t_{0}+\epsilon} \widehat{H}(t) dt\right)
= \exp\left(-ig\widehat{P}_{MD} \otimes \widehat{A} \right),
\end{equation}
where we have set
$\hbar=1$, and the interaction is only non-zero from $t_{0}-\epsilon$ to
$t_{0}+\epsilon$. The measurement interaction is weakened by minimizing $g/\sigma_{MD}$, where $g=\int_{+\infty}^{-\infty} g(t) dt$ is a constant defining the coupling strength, and $\sigma_{MD}$ is the uncertainty in the position observable for the MD. 

Let us imagine that the measurement device has an initial quantum state that is a real-valued Gaussian with unit uncertainty.  Then, in the limit of very weak coupling between the measuring device and the system, the weak measurement of the observable $\widehat{A}$ results in a shift in the measuring device's wavefunction so that
\begin{equation}\label{Eq:WeakValueShift}
\varphi(x) \rightarrow \varphi(x-g\re[A_w] ),
\end{equation}
where $g$ is the interaction coupling strength and $A_w$ is the weak value of the observable being measured.  
The weak value of an obervable 
$\widehat{A}$ conditioned on the pre-selected state $\rho_i$ and post-selected (pure) state $|\varphi\ra\la\varphi|$ is given by
\begin{equation}
^\varphi A_w^{\rho_i} = \frac{\Tr[|\varphi\ra\la\varphi| \widehat{A}\rho_i]}{\Tr[|\varphi\ra\la\varphi|\rho_i]}.
\end{equation}
Note that the weak value of a Hermitian observable can in general be complex valued, and even when real valued, it can lie outside the observable's eigenspectrum. 

If we condition our weak measurement results only on a particular (possibly mixed) initial state $\rho_i$, then the shift in the weak measurement device's wavefunction will be equal to the average of the weak values over all post-selected outcomes. Thus, the weak measurements will yield the expectation value $\la \widehat{A} \ra_{\rho_i}$ of $\widehat{A}$ at $\rho_i$. 
Intuitively, if a state was initialized as a pure state $\rho_i$, and then evolved to some other density operator $\rho_i^\p$ immediately prior to weak measurement of some observable $\wh{A}$, then we should expect the difference between the expectation value $\la \wh{A}\ra_{\rho_i}$ and the observed expectation value $\la\wh{A}\ra_{\rho_i^\p}$ to provide some insight into the evolution $\rho_i \mapsto \rho_i^\p$. In Section \ref{Sec:PE}, we will characterize the extent to which this intuition holds for the proposed QKD protocol.

\section{\label{Sec:WM-Protocol} Introducing a Weak Measurement Based Protocol}

In Table \ref{Tab:Protocol}, we lay out an example prepare-and-measure QKD protocol that utilizes weak measurements for parameter estimation. Note that, aside from the parameter estimation methods, the other striking difference between this protocol and BB84 is that the key is only generated from signals prepared and measured in the $Z$ basis. Obviously, we could construct an equivalent protocol in which Bob only measures in the $X$ basis. Consequently, if one augments the standard BB84 protocol with the above weak measurement steps, it is effectively equivalent to implementing the above protocol twice. Bob randomly choosing in which basis to measure can be viewed as an interleaving of the two protocols. We stress that, by utilizing the weak measurement results, the parameter estimation is no longer dependent on Bob's final basis choice. Consequently, neither randomness nor secrecy of Bob's basis choice need be assumed.

\begin{table}
\centering
\begin{tabular}{|c|l|}\hline
\  & \textbf{A WEAK MEASUREMENT QKD PROTOCOL} \\\hline\hline
\ & Alice generates a length $2n$ random string of bits $z \in \{0,1\}^{(2n)}$, encoding each bit in either the \\
1 & $X$ or $Z$ basis uniformly at random. Then she transmits each qubit to Bob.\\\hline
\ & Bob performs weak measurements of $\widehat{H}^+$ or $\widehat{H}^-$ chosen uniformly at random on each signal he \\ 
2 & receives, then he strongly measures in the $Z$ basis, recording both of his measurement results.\\\hline
\ & Bob openly shares the weak measurement results with Alice, who uses them to estimate the bit\\
3 & and phase error rates. If the error rates are too high, she aborts the protocol.\\\hline
\ & If the error rates are low enough, Alice announces which signals were prepared in the Z basis.\\
\ & Alice and Bob perform classical post-processing on these signals to correct the errors and distill\\
4 & a smaller, length $k$ secure key.\\\hline
\end{tabular}
\caption{Outline of a QKD protocol with weak measurements.}
\label{Tab:Protocol}
\end{table}

The degree to which a weak measurement interaction disturbs the signal is a direct function of the interaction coupling constant $g$. Because this is a controllable parameter, we can make it sufficiently small so as to be effectively negligible. Consequently, weakly measuring every signal has a negligible effect on the key rate. We will explicitly calculate the disturbance caused by the weak measurement interaction and its effect on the key rate in Section \ref{Sec:WM-disturbance}. 

\begin{figure}%
\centering 

\begin{tikzpicture}[scale=3]
\coordinate [] (origin) at (0,0);
\coordinate [label=right:$|+\ra\la+|$] (+) at (1,0);
\coordinate [label=left:$|-\ra\la -|$] (-) at (-1,0);
\coordinate [label=right:$\widehat{H}^+$] (H+) at (45:1cm);
\coordinate [label=above:$|0\ra\la 0|$] (0) at (0,1);
\coordinate [label=below:$|1\ra\la 1|$] (1) at (0,-1);
\coordinate [label=left:$\widehat{H}^-$] (H-) at (135:1cm);
\coordinate [label=left:$\widehat{H}^+_\bot$] (H+bot) at (-135:1cm);
\coordinate [label=right:$\widehat{H}^-_\bot$] (H-bot) at (-45:1cm);

\coordinate (noise) at ($ (0)!0.5!(+) $){};
\coordinate (securebound) at ($ (noise)!0.5!(H+) $){};


\node (+node) at (+) [circle, draw, fill, inner sep=1.5pt]{};
\node (H+node) at (H+) [circle, draw, fill, inner sep=1.5pt]{};
\node (-node) at (-) [circle, draw, fill, inner sep=1.5pt]{};
\node (0node) at (0) [circle, draw, fill, inner sep=1.5pt]{};
\node (1node) at (1) [circle, draw, fill, inner sep=1.5pt]{};
\node (H-node) at (H-) [circle, draw, fill, inner sep=1.5pt]{};
\node (H+botnode) at (H+bot) [circle, draw, fill, inner sep=1.5pt]{};
\node (H-botnode) at (H-bot) [circle, draw, fill, inner sep=1.5pt]{};

\node (Circ) [draw,circle through=(H+)] at (origin) {};

\draw (0)--(1);
\draw (-)--(+);

\end{tikzpicture}%
\caption{A view of the geometric relationship between the $\wh{H}^\pm$ projectors, their complements $\wh{H}^\pm_\bot$, and the projectors onto the four states used in BB84.}\label{Fig:CirclePlot}%
\end{figure}
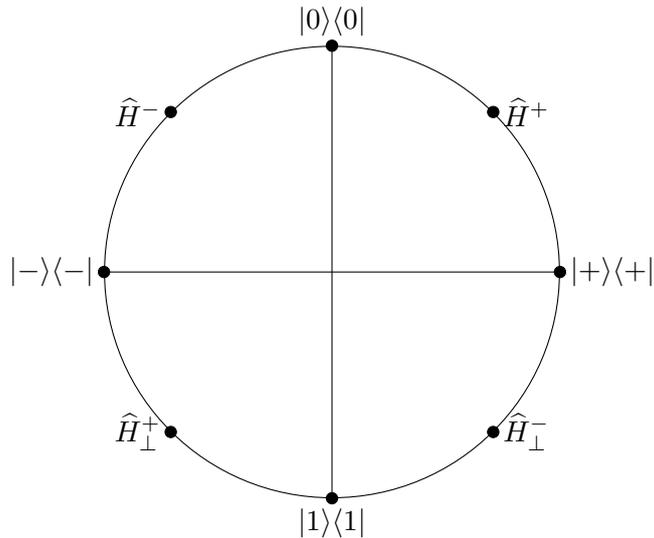

With respect to the weak measurements, we will be particularly interested in the two observables\footnote{
Another pair of weak measurement observables that could be used to provide the same information as $\wh{H}^\pm$ are the trace-0 observables $\wh{S}^\pm = \wh{H}^\pm - \wh{H}^{\pm}_\bot = \frac{1}{\sqrt{2}}(\wh{Z} \pm \wh{X})$. For concreteness, however, we will simply focus on $\wh{H}^\pm$.
}
\begin{equation}\label{Eq:H}
\widehat{H}^\pm \equiv \frac{1}{2}\left[ \widehat{I} + \frac{1}{\sqrt{2}}(\widehat{Z} \pm \widehat{X})\right].
\end{equation}
%
Figure \ref{Fig:CirclePlot} demonstrates the geometric relationship between the above observables, their complements $\wh{H}^\pm_\bot$, and the projectors onto the four states used in the standard BB84 protocol in the density operator framework. While in principle it is possible for Bob to measure both $\wh{H}^+$ and $\wh{H}^-$ on each signal, it is practically easier and, as we will see in Section \ref{Sec:MDI-security}, advantageous for the protocol's security for Bob to randomly choose one of the observables to measure on each signal. 

\section{\label{Sec:PE} Parameter Estimation with Weak Measurements}

Lo and Chau \cite{LoChau} were the first to rigorously relate the security of QKD against arbitrary attacks to entanglement distillation. In so doing, they demonstrated that the ability of Alice and Bob to distill a secure key from their correlated signals is precisely a function of how coupled their signals are to the environment. Shor and Preskill \cite{ShorPreskill} showed that entanglement distillation with one-way communication was equivalent to quantum error correcting codes. As a consequence of the fact that $X$ errors are decoupled from $Z$ errors in quantum error correcting codes, determining the $X$ and $Z$ error rates on the signals suffices to determine the coupling of the signal with the environment. 

Gottesman, Lo, L\"{u}tkenhaus, and Preskill (GLLP) \cite{GLLP} further observed that the above holds under the assumption that the coupling is basis independent. In that case, the bit error rate, $\delta_b$, directly measured from the signal $X$ and $Z$ errors, is approximately equal to the phase error rate, $\delta_p$. Since $\delta_p$ cannot be directly measured, it must be inferred. If Eve can leverage even a small amount of basis information, possibly through device imperfections, then equality cannot be assumed. However, if it is possible to bound this leakage, then GLLP demonstrated that it was often possible to place an upper bound on how much larger than $\delta_b$ the phase error rate could be, and consequently security could be preserved.

The general procedure for proving asymptotic security of any prepare-and-measure (PAM) QKD protocol is to design an equivalent entanglement distillation (ED) protocol. Once the ED protocol can be proven secure, the corresponding security of the PAM protocol follows. In that vein, we will first characterize an equivalent weak measurement ED protocol. Aside from the weak measurement parameter estimation step, there is no fundamental difference between this ED protocol and the ones described in, e.g. \cite{LoChau,ShorPreskill, GLLP}. In these asymptotic security proofs, a fundamental assumption is that the bit error rate, $\delta_b$, can be accurately estimated during parameter estimation. Consequently, as long as we can demonstrate that the weak measurement parameter estimation step is accurate, then we have validated this assumption, and security follows.

\subsection{\label{Subsec:ED} Weak Measurement Entanglement Distillation Protocol}

Consider the following entanglement distillation protocol:
\begin{itemize}
\item[1)]	Alice creates $n+m$ pairs of maximally entangled qubits
\begin{equation}
|\phi^+\ra^{\otimes (n+m)} = \frac{1}{\sqrt{2}}(|00\ra + |11\ra)^{\otimes (n+m)},
\end{equation}
and sends half of each pair to Bob.
\item[2)] Alice selects a random subset of $m$ of these signals to sacrifice as ``check bits.'' On these signals, Alice randomly measures $X$ or $Z$ on her half of the entangled pair, while Bob performs weak measurements on his half of each the entangled pairs and broadcasts his weak measurement results. Alice estimates the error rates $\delta_X$ and $\delta_Z$. If the error rates are too high, they abort the protocol.
\item[3)]	Assuming the error rates are below some predefined threshold, Alice and Bob perform entanglement distillation on the remaining $n$ signals, extracting a smaller number $k \leq n$ of high-fidelity entangled pairs from the $n$ noisy ones.
\item[4)]	Alice and Bob both measure $Z$ on each of their $k$ pairs, producing a $k$-bit shared random key about which Eve has negligible information.
\end{itemize}

If the ED protocol has one-way communication, then it can be shown to be equivalent to quantum error correcting codes \cite{ShorPreskill}. In particular, using CSS codes, the bit errors are decoupled from the phase errors. Now, the reason for estimating both the bit and phase errors in the transmitted signals is so that we may estimate how strongly coupled Eve's probe state is to the signal state. Because the key is determined from the subset of signals for which Alice and Bob both measure in the same basis, they only need to correct the bit errors. Since this can be done using strictly classical error correction, Alice and Bob may just as well measure $Z$ on the $n$ noisy pairs, and then reserve error correction for the classical post-processing step. Furthermore, extracting $k$ secure bits from the $n$ shared (and now corrected) bits can be done using classical hashing techniques \cite{LoChau}. Since Alice and Bob are no longer worried about waiting until after entanglement distillation to measure their states, they may as well use the weak measurement results of every signal to perform parameter estimation, reserving only the subset of signals for which Alice and Bob's bases agree for key generation. In this way, the PAM protocol is equivalent to the ED protocol. Consequently, the security of the PAM protocol follows from the security of the ED protocol.

In order to obtain a secure key, then, it is imperative that Alice and Bob obtain accurate estimates of both the $X$ and $Z$ errors, which we denote $\delta_X$ and $\delta_Z$, respectively. In the standard approach to parameter estimation, Alice and Bob sacrifice a random subset of signals prepared in each basis to be used as ``check bits,'' from which they may obtain decent estimates of these two parameters as long as Bob measures in the same basis each check bit signal was prepared. Now, if Eve were to know in advance which signals were going to be used as check bits, then she could avoid interacting with those signals altogether so that Alice and Bob's estimates would not be correlated with the actual error parameters on the remaining signals.

In order to ensure Eve does not know in advance which signals are check bits, Alice keeps this information secret until after Bob receives her signals. Ideally, Bob would want to keep his signals in a quantum memory until after Alice announces which signals are check bits. Then, Alice can tell Bob in which basis to measure each check bit. Assuming the error rates are sufficiently low, it is now safe for Alice to reveal the basis information on the remaining signals to Bob, so that he can measure accordingly.

In practice, however, Bob has no quantum memory. In prepare-and-measure protocols, the work-around is for Bob to measure each signal as it arrives, and effectively guess the preparation basis. If his guess is a basis different from the one in which the state was prepared, then his measured result is completely uncorrelated with the signal Alice transmitted. Consequently, these signals can be used neither for key generation nor for parameter estimation, so they are removed from each list.

The last point is of critical importance -- in the standard approach to parameter estimation, Bob must measure in the correct basis in order for his measurement results to reveal any information about the channel parameters. For this reason, it is impossible for the standard method of parameter estimation to be independent of Bob's measurements. This critical dependence has been utilized repeatedly to hack QKD protocols \cite{Lyd10,Gerh11,Sajeed16,Makarov16}. By taking control over Bob's detectors, Eve can assure that the signals that could best detect her interactions will never be used for parameter estimation.\footnote{While the parameters estimated in a successfully implemented intercept-resend attack with detector blinding will accurately reflect the error estimates along the subset of received signals to be used for key generation, they are not an accurate characterization of the error parameters over the entire collection of signals that Bob received.} This weakness has led to the development of Measurement-Device-Independent QKD (MDI-QKD) protocols \cite{LoCurQi12,Bing13,Curty14,Tang14,Tittel15} that eliminate Bob's detectors altogether. To achieve this, MDI-QKD protocols perform parameter estimation and key generation based on the results of a joint Bell measurement on both Alice and Bob's signals. This Bell state measurement is probabilistically implemented by two-photon interference at the receiver. By replacing Bob's detectors with an additional source, Eve can only leverage device imperfections that leak information to optimize her attacks, rather than altogether controlling Bob's basis choices.

Weak measurements, on the other hand, provide a fundamentally new avenue for parameter estimation that, like MDI-QKD, \emph{is not dependent on Bob's final measurements}. Unlike the standard approach to parameter estimation, a weak measurement does not appreciably disturb the signal. Thus, rather than sacrificing a small percentage of the total number of received signals for parameter estimation, we may sacrifice a small percentage of \emph{each} signal for parameter estimation. Because the error information is obtained from the weak measurement results and not Bob's final measurement, what Bob does with these signals afterwards does not matter from a parameter estimation perspective. Consequently, Bob may as well strongly measure every signal as it arrives, and furthermore may measure all of them in the same basis. 

The security benefits of weak measurements are now apparent. In the standard method of parameter estimation, if Eve has complete control of Bob's detectors, she can enforce that certain of her interactions with the signal will be removed from the parameter estimation step. With weak measurements of the $\widehat{H}$ projectors, on the other hand, this line of attack is completely avoided by the fact that \emph{every} signal Bob receives contributes an equal amount of information for parameter estimation.

\subsection{\label{Subsec:PE} Estimating the Error Parameters from Expectation Values}

While it is possible in principle to weakly measure both observables on each signal, for practical reasons and, as we shall see in Section \ref{Sec:MDI-security}, to maximize the protocol's security, we will suppose that immediately prior to Bob's final measurement in the $Z$ basis, he randomly chooses one of the $\wh{H}$ observables from Eq. \eqref{Eq:H} to weakly measure on the arriving signal, and records his results. By averaging these measurement results over many signals, and then conditioning on each of the four possible initial states, Alice and Bob can obtain precise estimates of the expectation values for each of the $\wh{H}$ observables at each of the initial states. 

Using the Bloch representation, an arbitrary qubit density operator $\rho$ can be decomposed as
\begin{equation}\label{Eq:rho}
\rho = \frac{1}{2}\left(\mbb{I} + r_xX + r_yY + r_zZ\right),
\end{equation}
for some $r_x, r_y, r_z \in [-1, 1]$.
The expectation value of $\wh{H}^\pm$ at $\rho$ is then given by
\begin{equation}
\la \wh{H}^\pm \ra_\rho = \Tr(\wh{H}^\pm \rho) = \frac{1}{2}\left(1 + \frac{r_z \pm r_x}{\sqrt{2}}\right).
\end{equation}
From this, we obtain
\begin{align} \label{Eq:r_x}
r_x &= \sqrt{2}(\la \wh{H}^+\ra_\rho - \la \wh{H}^-\ra_\rho)\\ \label{Eq:r_z}
r_z &= \sqrt{2}(\la \wh{H}^+\ra_\rho + \la \wh{H}^-\ra_\rho - 1).
\end{align}

Now from $r_x$ and $r_z$ we may arrive at the error parameters $\delta_X$ and $\delta_Z$. More explicitly, let 
\begin{equation}\label{Eq:rho-alpha}
\rho_\alpha := \frac{1}{2}\left(\mbb{I} + r_x^\alpha X + r_y^\alpha Y + r_z^\alpha Z\right)
\end{equation}
denote the density operator that characterizes the state of a signal initially prepared as $|\alpha\ra\la\alpha |$ after being subjected to the quantum channel, where $\alpha \in \{0,1,+,-\}$. Under the assumption that each signal was transmitted with equal probability, it follows that
\begin{align}
\delta_X &= \frac{1}{4}(2- r_x^+ + r_x^-),\\
\delta_Z &= \frac{1}{4}(2- r_z^0 + r_z^1).
\end{align}
If the qubit channel is unital, it follows that $r_x^+ = -r_x^-$ and $r_z^0 = -r_z^1$, so that the above simplifies to
\begin{align}\label{Eq:deltaX}
\delta_X &= \frac{1-r_x^+}{2} = \frac{1+r_x^-}{2},\\ \label{Eq:deltaZ}
\delta_Z &= \frac{1-r_z^0}{2} = \frac{1+r_z^1}{2}.
\end{align}
Furthermore, we have
\begin{equation}\label{Eq:H-unity}
\la \wh{H}^\pm \ra_{\rho_\alpha} + \la \wh{H}^\pm \ra_{\rho_{\alpha_{\bot}}} = 1,
\end{equation}
where we define $|\alpha_\bot\ra$ to be the orthogonal complement of $|\alpha\ra$, for $\alpha \in \{0,1,+,-\}$.

Suppose the coupling strengths of the weak measurement of $\wh{H}^\pm$  are given by $g^\pm$, respectively. Then conditioning on the subset of signals prepared in the state $|\alpha\ra\la\alpha|$, the averages of the weak measurement results tend toward the values
\begin{align}
\mu^+_\alpha &:= g^+\la \wh{H}^+\ra_{\rho_\alpha}\\
\mu^-_\alpha &:= g^-\la \wh{H}^-\ra_{\rho_\alpha}.
\end{align}
Consequently, by Eq. \eqref{Eq:H-unity} we obtain
\begin{align}\label{Eq:g+}
g^+ &= \mu^+_\alpha + \mu^+_{\alpha_{\bot}}\\\label{Eq:g-}
g^- &= \mu^-_\alpha + \mu^-_{\alpha_{\bot}}.
\end{align}

Thus, the coupling parameters need not be known in advance. That is to say, we may allow for the possibility that Eve secretly controls the coupling parameter, or that it may change over time due to hardware imperfections, and estimate the parameter separately in each round of key generation.

After Bob receives sufficiently many signals, he publicly broadcasts the choice of weak measurement observable and corresponding weak measurement results for each signal he receives. Because Alice is the only individual who knows which signals were transmitted, she is the only one who can condition the weak measurement results according to the initial states. She uses the weak measurement results of every signal Bob receives in order to perform parameter estimation accordingly.

A natural question to ask is why we have chosen to use the two observables $\wh{H}^\pm$. Indeed, one may envision a potentially far simpler weak measurement based protocol in which Bob weakly measures $\wh{X}$ on each signal, and strongly measures $\wh{Z}$. In this version, $\delta_Z$ could be calculated in the traditional way using a sample of signals prepared and measured in the $Z$ basis, while $\delta_X$ could be estimated from the weak measurement results of signals prepared in the $X$ basis using methods similar to that above. Indeed, such an approach would work well under the assumption that the weak measurements were implemented perfectly. However, if there are device imperfections, or if Eve has any control over the weak measurement interactions, then this example no longer becomes secure. The projectors $\wh{H}^+$ and $\wh{H}^-$ were chosen precisely because they provide the same amount of information about both the $X$ and $Z$ bases. This allows the protocol to extract an equal amount of information about \emph{both} error rates for every signal received.

We remark here that GLLP used as their bit error rate $\delta_b \approx \frac{1}{2}(\delta_X + \delta_Z)$. Because key is only extracted from signals prepared and measured in the $Z$ basis in the weak measurement protocol, one may be na{\"i}vely compelled to assign $\delta_b = \delta_Z$ and $\delta_p = \delta_X$ in the weak measurement protocol. However, it is only the case that $\delta_p = \delta_X$ under the assumption that Eve's attack is basis independent, so that the bit-flip errors on signals prepared in the $X$ basis are representative of the phase-flip errors on the signals prepared in the $Z$ basis. Moreover, if this assumption of basis independence is true, then it necessarily follows that $\delta_p \approx \delta_b$, in which case, $\delta_b \approx \frac{1}{2}(\delta_X + \delta_Z)$ trivially holds. By the concavity of entropy, the following necessarily holds for all $\delta_X$ and $\delta_Z$:
\begin{equation}
1 - 2H_2(\delta_b) \leq 1-H_2(\delta_X)-H_2(\delta_Z),
\end{equation}
where $H_2$ is the binary entropy function. Hence, there is no reduction in security by choosing $\delta_b$ as our estimate of the bit error rate, even if the assumption of basis independence does not hold. Furthermore, as we shall see, adversarial attempts to leverage weak measurement device imperfections to decrease the estimate of $\delta_X$ will generally increase the estimate of $\delta_Z$ and conversely. We will analyze the effect of weak measurement imperfections on the protocol's error estimates in Section \ref{Sec:WM-Security}.


\section{Effect of Weak Measurement Disturbance on Key Rate}\label{Sec:WM-disturbance}

Since even weak measurements will disturb the qubits being measured, the actual quantum bit error rate (QBER) will be somewhat larger than what is estimated from the weak measurement results. In this section we will calculate this additional error rate and show it has a very small effect on the secure key rate. 

Since we are focusing only on the contribution to $\delta_X$ and $\delta_Z$ due to the weak measurement back-action, we assume that the state just before Bob's weak measurement is pure.  Let that state be the initial qubit state $|\psi\ra = \alpha|H_\bot\ra + \beta|H\ra$ written in the $H$ basis.  Then, just after the interaction implementing the weak measurement of $|H\ra\la H|$, using a Gaussian measurement device (MD) wavefunction with variance $\sigma^2$ for the weak measurement pointer state, we have the following for the MD and qubit combined system,
\begin{align}
 \rho &= |\alpha|^2|H_\bot \ra\la H_\bot | \left(\frac{1}{\sigma\sqrt{2 \pi}}\right) \int \! e^{-x^2/4 \sigma^2} \int \! e^{-y^2/4 \sigma^2}|x\ra\la y| \mathrm{d}x \mathrm{d}y \nonumber\\
 \ &+  |\beta|^2|H \ra\la H | \left(\frac{1}{\sigma\sqrt{2 \pi}}\right) \int \! e^{-(x-g)^2/4 \sigma^2} \int \! e^{-(y-g)^2/4 \sigma^2}|x\ra\la y| \mathrm{d}x \mathrm{d}y \nonumber \\ 
 \ &+  \alpha\beta^*|H_\bot \ra\la H | \left(\frac{1}{\sigma\sqrt{2 \pi}}\right) \int \! e^{-x^2/4 \sigma^2} \int \! e^{-(y-g)^2/4 \sigma^2}|x\ra\la y| \mathrm{d}x \mathrm{d}y \nonumber \\
 \ &+ \alpha^*\beta|H \ra\la H_\bot | \left(\frac{1}{\sigma\sqrt{2 \pi}}\right) \int \! e^{-(x-g)^2/4 \sigma^2} \int \! e^{-y^2/4 \sigma^2}|x\ra\la y| \mathrm{d}x \mathrm{d}y.
\end{align}
Tracing out the MD gives the density operator for the qubit after the weak measurement interaction,
\begin{align}
 \rho_{qubit} &=\mathrm{Tr}_{MD}(\rho)=\int \! \la x|\rho|x\ra \mathrm{d}x \nonumber  \\ 
&= |\alpha|^2|H_\bot \ra\la H_\bot| + |\beta|^2|H \ra\la H | + e^{-g^2/8 \sigma^2} \left( \alpha\beta^*|H_\bot \ra\la H | + \alpha^*\beta|H \ra\la H_\bot | \right) \nonumber \\ 
&=  e^{-g^2/8 \sigma^2}|\psi\ra\la \psi| + \left(1- e^{-g^2/8 \sigma^2}\right)\left( |\alpha|^2|H_\bot \ra\la H_\bot| + |\beta|^2|H \ra\la H | \right).
\end{align}
The last expression shows that the qubit density operator is a mixture of the undisturbed original state and collapses onto each of the two $\widehat{H}$ basis states. Using the BB84 input states, it can be easily shown from this that the probability for the weak measurement to collapse the initial qubit state into its orthogonal state is 
\begin{equation}\label{Eq:WM_error}
\delta_{wm} = \frac{1}{4} \left[ 1 - \exp \left( - \frac{g^2}{8 \sigma^2} \right) \right]
\end{equation}
for all four initial qubit states and for the weak measurement of either $\wh{H}^\pm$ projector.  This means that the weak measurement disturbance acts as a depolarizing channel with an error probability given by $\delta_{wm}$, which must be added to the estimate of the QBER.  To verify that the weak measurements are indeed weak, as well as to accurately estimate the weak measurement induced error rate, it is important to estimate the variance of the weak measurement results, $\text{Var}[\mu_\alpha^\pm]$, in addition to $g^\pm$. As we will see in Section \ref{Sec:MDI-security}, estimating $\text{Var}[\mu_\alpha^\pm]$ will also provide a very strong defense against Eve successfully breaking the protocol if we grant Eve the ability to directly control the weak measurement results. While we have assumed that the weak measurement pointer was a real valued Gaussian wavefunction in the analysis above, very similar results hold for any MD wavefunctions that are real valued and symmetric, i.e. $|\varphi(x)| = |\varphi(-x)|$, as shown by Silva, \emph{et al}. \cite{Silva15}. 

\begin{figure}
\centering
\includegraphics[width=0.5\textwidth]{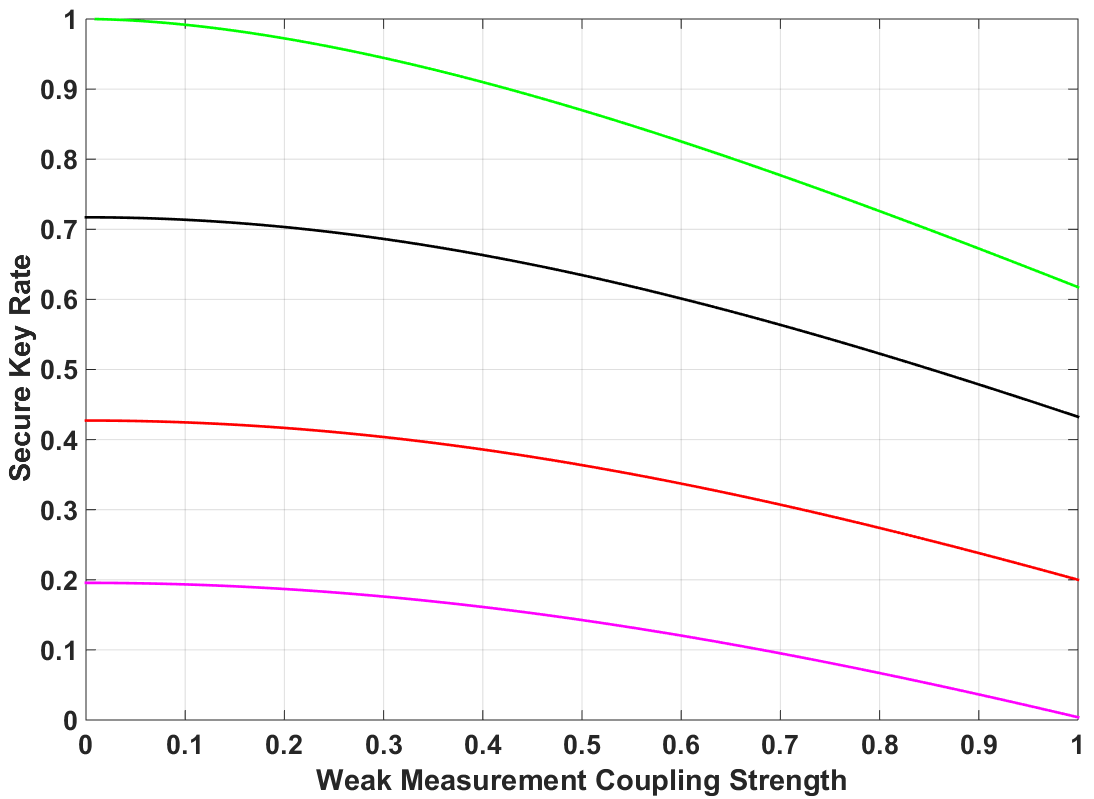}
\caption{The above graphs show the reduction in secure key rate as a function of the coupling strength ($g/\sigma_{MD}$) when the signals are subjected to depolarizing noise with a channel error rate of 0\% (Green), 2\% (Black), 5\% (Red), and 8\% (Magenta). For the sake of simplicity, key rates $R$ were calculated here using the idealized $R = 1-2H_2(QBER)$, where $H_2$ is the binary Shannon entropy, and the QBER is the sum of the channel noise and $\delta_{wm}$.}\label{Fig:SecKey_coupling}%
\end{figure}

For a practical quantum optical weak measurement, the ratio of the coupling strength and MD pointer position uncertainty can be less than $0.10$.  For weak measurements of this strength, the measurement induced error is $\delta_{wm} < 0.0004$.  Thus, even for practical, non-zero strength weak measurements, the induced error rate from the weak measurements is less than $0.1\%$. Figure \ref{Fig:SecKey_coupling} graphs the reduction in key rate as a function of the coupling strength $g/\sigma_{MD}$ when the signals are subjected to varying degrees of depolarizing noise. In particular, we see that in an otherwise noiseless quantum channel, setting the coupling strength to $0.10$ will reduce the overall key rate by less than one percent, becoming much less than this when the coupling strength is no longer the dominant source of noise.

\section{\label{Sec:WM-Security} Security of Parameter Estimation}

As previously explained, in order to estimate Eve's coupling to the quantum signal, we must obtain accurate estimates of both the $X$ and $Z$ error rates on the signal. Consequently, the security of the protocol can be reduced to an analysis of the accuracy of the parameter estimation step in an adversarial environment. In the ED protocol, there is an implicit assumption that Eve's coupling with the signals is independent of whether or not a given signal is used for parameter estimation. Under this assumption, the parameter estimates performed on the subset of signals used as check bits will be an accurate reflection of the parameters on the full collection of signals. This assumption is not needed in the weak measurement PAM protocol, as \emph{every} signal Bob receives contributes some information towards parameter estimation. Consequently, we may assume without any loss of generality, or any reduction in security, that Eve cannot condition her interactions with the signals based on whether or not they are used for parameter estimation.

The security analysis in the various scenarios will all take the following assumption as given:
\begin{assumption}\label{Asm:PathManipulation}
Eve cannot manipulate the signal in the interval between the weak measurement interaction and Bob's final measurement.
\end{assumption}
The parameters estimated from the weak measurements are assumed to characterize properties of the signal leading up to the moment of weak measurement. It is precisely this assumption that allows us to decouple the parameter estimation from its dependence on Bob's detectors. This assumption can be physically enforced by making these weak measurement interactions immediately prior to Bob's strong measurement, so that the path between the weak and strong measurements is very small and is entirely contained inside of a secured portion of Bob's signal measurement device. This is essentially the assumption that Eve has not covertly placed any qubit altering components or measuring apparatus inside this small, secured section of Bob's measurement device. 

Security analysis under the assumption of accurate parameter estimation has been thoroughly investigated in, e.g. \cite{LoChau,ShorPreskill, GLLP}, and we do not wish to reproduce their results here. Since their results are immediately applicable to the weak measurement QKD protocol as long as the weak measurement parameter estimations are accurate, our security analysis will be reduced to an analysis of the accuracy of the parameter estimation step. To be explicit, we make the following remark:

\begin{remark}\label{Rem:SecPreserve}
The security of the weak measurement QKD protocol is considered \emph{preserved}, with asymptotic security following from, e.g. \cite{LoChau,ShorPreskill, GLLP}, if the estimated quantum bit error rate is either above the security threshold or at least as large as the actual quantum bit error rate.
\end{remark}

The following is an immediate consequence:
\begin{theorem}\label{Thm:PerfectWM}
Under the assumption that Bob implements his weak measurements perfectly, leaks no information about his choice of weak measurement observable, and Eve cannot manipulate the weak measurement observables, then security of the weak measurement QKD protocol is preserved.
\end{theorem}
In other words, if weak measurement parameter estimation is idealized, then beyond the fact that Eve can no longer implement any detector basis-dependent attacks, the asymptotic security analysis is no different than that of BB84.

We now consider various scenarios with practical imperfections in the implementation of the protocol, and verify that security is preserved, as in Remark \ref{Rem:SecPreserve}, in each one. As will be demonstrated, device imperfections may sometimes cause the parameter estimates to differ from the actual parameter values. For notational clarity, we will distinguish an estimate of a parameter determined by the weak measurement results from the actual parameter value by placing a tilde over the parameter.

\subsection{Imperfect Weak Measurements}\label{Subsec:ImpWM}
\begin{figure}%
\centering 

\begin{tikzpicture}[scale=3]
\coordinate [] (origin) at (0,0);
\coordinate [] (+) at (1,0);
\coordinate [] (-) at (-1,0);
\coordinate [label=right:$\widehat{H}^+$] (H+) at (45:1cm);
\coordinate [] (H+theta+) at (55:1cm);
\coordinate [] (H+theta-) at (35:1cm);
\coordinate [] (0) at (0,1);
\coordinate [] (1) at (0,-1);
\coordinate [label=left:$\widehat{H}^-$] (H-) at (135:1cm);
\coordinate [] (H-theta+) at (145:1cm);
\coordinate [] (H-theta-) at (125:1cm);

\node (H+node) at (H+) [circle, draw, fill, inner sep=1.5pt]{};
\node (H-node) at (H-) [circle, draw, fill, inner sep=1.5pt]{};

\node (Circ) [draw,circle through=(H+)] at (origin) {};

\coordinate [label=above right:$\theta$] (theta) at (35:0.5cm); 

\draw (0)--(1);
\draw (-)--(+);
\draw[color=gray] (origin)-.(H+theta+);
\draw (origin)--(H+);
\draw[color=gray] (origin)-.(H+theta-);

\fill[fill=gray, opacity=.4] (origin)--(H+theta-) arc (35:55:1cm)--(origin); 

\draw (origin)--(H-);
\draw[color=gray] (origin)-.(H-theta+);
\draw[color=gray] (origin)-.(H-theta-);
\fill[fill=gray, opacity=.4] (origin)--(H-theta-) arc (125:145:1cm)--(origin);

\draw (theta) arc (35:45:0.5cm);

\end{tikzpicture}%
\caption{If the weak measurement interactions are imprecise, then the observable weakly measured on any given signal may actually be another projector ``near'' the correct one.}\label{Fig:NoisyWM}%
\end{figure}
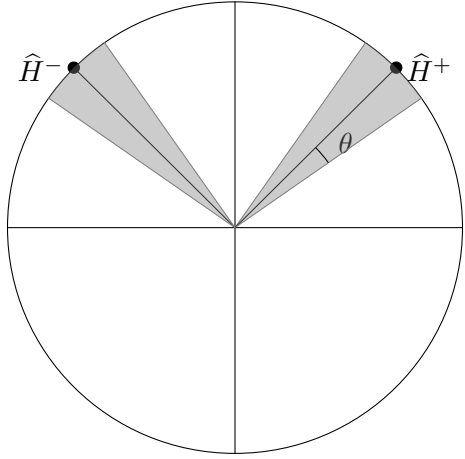

Let us suppose that, when Bob performs a weak measurement interaction, there is a small degree of uncertainty with respect to which observable he is weakly measuring. More specifically, let us suppose that for any given signal, the observable Bob actually measures is a projector that lies within some cone of half-angle $\theta$ around the correct projector (see Figure \ref{Fig:NoisyWM}).

In this case, we should expect the average of the weak measurement results to be given by
\begin{equation}
\mu_\alpha^\pm = g^\pm \int_{-\theta}^{\theta} \la \wh{H}^\pm (\varphi) \ra_\alpha \ \text{Prob}(\varphi) \ d\varphi
\end{equation}
where Prob$(\varphi)$ is the probability distribution for $\varphi$ and $\wh{H}^\pm (\varphi)$ is the projector given by 
\begin{equation}\label{Eq:NoisyH}
\wh{H}^\pm(\varphi) := \frac{1}{2}\left(\mbb{I} \pm \sin\left(\frac{\pi}{4} + \varphi\right)X + \cos\left(\frac{\pi}{4} + \varphi\right)Z\right).
\end{equation}
If we further assume that the noise applied to the projector's angle is additive Gaussian noise around the intended angle with variance $\sigma_\varphi^2$, then the estimated mean pointer shift is unchanged, 
\begin{equation}
\mu_\alpha^\pm =\ g^\pm \la \wh{H}^\pm \ra_\alpha,
\end{equation}
while the estimated variance of the pointer shift will have an additional component that is defined by the imprecision of $\varphi$. For small degrees of imprecision, with $\sigma_\varphi^2 \ll 1$, the variance in the weak measurement pointer results are now given by



\begin{equation}
\text{Var}[\mu_\alpha^\pm] = (g^\pm)^2 \left(\sigma_{MD}^2 + \frac{1}{4} \sigma_\varphi^2\right).
\end{equation}
Thus in the limit of many signals, the average weak measurement results will be unaffected, and $\delta_X$ and $\delta_Z$ are unchanged. 

However, the estimated variance of the weak measurements will increase due to the imprecision of the angle determining the projectors. Because Alice and Bob use the estimated variance to place an upper bound on the weak measurement induced bit error rate, $\delta_{wm}$, they will need to place an upper bound, $\sigma_\varphi^U$, on the amount of permissible noise in the weak measurement interaction implementation. Using this upper bound on the implementation inaccuracy, Alice and Bob will estimate the lower bound on the weak measurement pointer variance using
\begin{equation}\label{Eq:WM-Var}
\sigma_{MD}^2 \leq \frac{\text{Var}[\mu_\alpha^\pm]}{g^2} - \frac{1}{4}\left(\sigma_{\varphi}^U\right)^2.
\end{equation}
If Alice and Bob overestimate $\sigma_{MD}$ then their estimate $\tilde{\delta}_{wm}$ of the weak measurement induced error rate will be smaller than the actual induced error rate $\delta_{wm}$. However, with an appropriate safety margin built into $\sigma_\varphi^U$, the protocol will work properly. If the noise in the angle is stationary and accurately characterized, then in the limit of infinitely long key, the bound on $\sigma_{MD}$ can be made tight and the secure key rate is unchanged. Thus we have shown:

\begin{theorem}
Suppose that Bob's implemented weak measurement projectors range over a small bounded region, with a Gaussian distribution symmetric about the correct projectors, but Eve otherwise has no direct control over the weak measurements. Then the security of the weak measurement QKD protocol is preserved.
\end{theorem}

\subsection{Biased Weak Measurements}

\begin{figure}%
\centering 

\begin{tikzpicture}[scale=3]
\coordinate [] (origin) at (0,0);
\coordinate [] (+) at (1,0);
\coordinate [] (-) at (-1,0);
\coordinate [label=above right:$\widehat{H}^+$] (H+) at (45:1cm);
\coordinate [label=above:$\widehat{H}^+(\varphi)$] (H+theta-) at (60:1cm);
\coordinate [] (0) at (0,1);
\coordinate [] (1) at (0,-1);
\coordinate [label=above left:$\widehat{H}^-$] (H-) at (135:1cm);
\coordinate [label=left:$\widehat{H}^-(\varphi)$] (H-theta-) at (150:1cm);

\node (H+node) at (H+) [circle, draw, fill, inner sep=1.5pt]{};
\node (H-node) at (H-) [circle, draw, fill, inner sep=1.5pt]{};
\node (H_varphi) at (H+theta-) [circle, draw, fill, inner sep=1.5pt]{};
\node (H_varphi) at (H-theta-) [circle, draw, fill, inner sep=1.5pt]{};

\node (Circ) [draw,circle through=(H+)] at (origin) {};

\coordinate [label=right:$\varphi$] (varphi) at (60:0.5cm); 
\coordinate [label=above:$\varphi$] (varphi2) at (150:0.5cm);

\draw (0)--(1);
\draw (-)--(+);
\draw (origin)--(H+theta-);
\draw (origin)--(H+);
\draw (origin)--(H-);
\draw (origin)--(H-theta-);
\draw (varphi) arc (60:45:0.5cm);
\draw (varphi2) arc (150:135:0.5cm);

\end{tikzpicture}%
\caption{A depiction of biased weak measurement observables.}\label{Fig:BiasedWM}%
\end{figure}
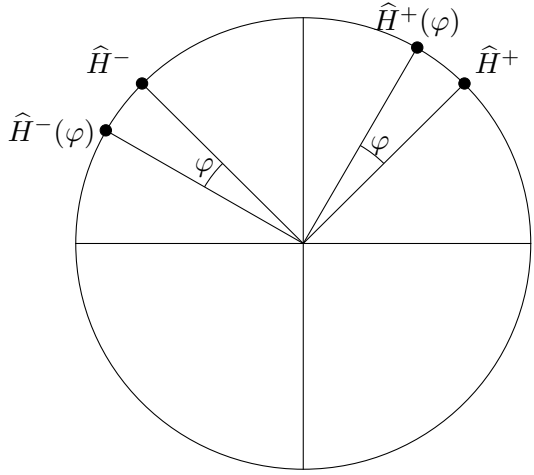

In the above, we considered the scenario in which the weak measurements were imperfect, yet unbiased. We will now consider what happens when the implemented weak measurement observables are biased by a fixed amount away from the correct observables in one direction. In other words, let us suppose that Eve has some method of rotating the weak measurement interaction, unbeknownst to Alice or Bob (see, e.g. Figure \ref{Fig:BiasedWM}). Even in this strongly adversarial environment, it is still nevertheless possible to obtain nonzero secure key. In fact, we will demonstrate that any rotation of both of the weak measurement observables will only serve to increase, rather than decrease, the estimated error, and hence preserve security.

More precisely, let us suppose that, whenever Bob intends to weakly measure $\wh{H}^\pm$, what he actually weakly measures is $\wh{H}^\pm(\varphi)$, for some fixed $\varphi \in (-\pi, \pi)$, where $\wh{H}^\pm(\varphi)$ is defined as in Eq. \eqref{Eq:NoisyH}. Then the expectation values $\la \wh{H}^\pm(\varphi) \ra_{\rho_\alpha}$ change according to:
\begin{equation}
\la \wh{H}^\pm(\varphi) \ra_{\rho_\alpha} = \frac{1}{2}\left( 1 \pm r_x^\alpha \sin\left(\frac{\pi}{4} + \varphi\right) + r_z^\alpha \cos\left(\frac{\pi}{4}+\varphi\right)\right).
\end{equation}

If the actual error rates are given by Eqs. \eqref{Eq:deltaX} and \eqref{Eq:deltaZ}, then using Eqs. \eqref{Eq:r_x} and \eqref{Eq:r_z}, the calculated error rates will be given by
\begin{align}
\tilde{\delta}_X &= \frac{1}{2} -\frac{r_x^+}{2}\left(\cos\left(\varphi\right) + \sin\left(\varphi\right)\right)\\
\tilde{\delta}_Z &= \frac{1}{2} -\frac{r_z^0}{2}\left(\cos\left(\varphi\right) - \sin\left(\varphi\right)\right).
\end{align}

\begin{figure}
\centering
    \begin{subfigure}[b]{0.5\textwidth}
    		\centering
                \includegraphics[width=0.95\linewidth]{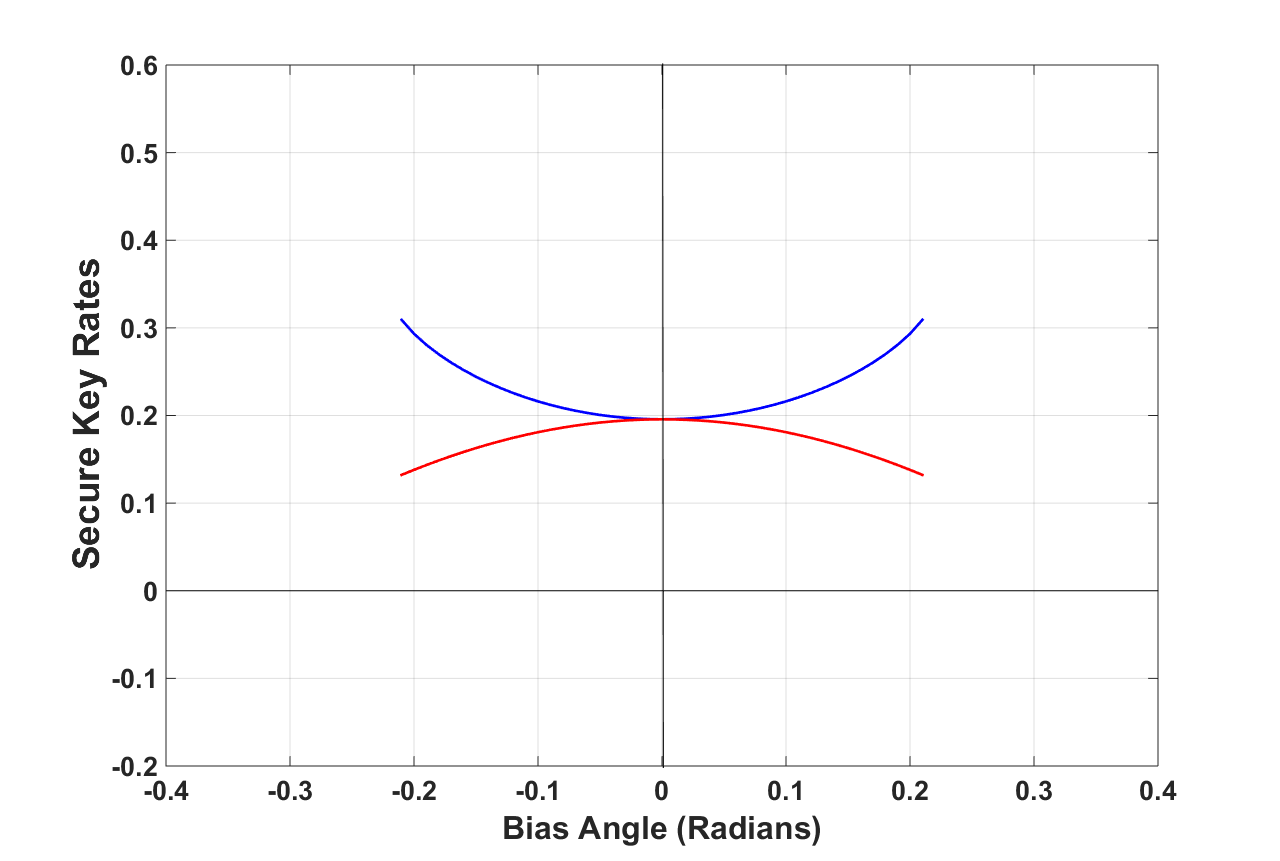}
                \caption{}\label{SubFig:QBER-08}%
        \end{subfigure}%
    \begin{subfigure}[b]{0.5\textwidth}
        	\centering
                \includegraphics[width=0.95\linewidth]{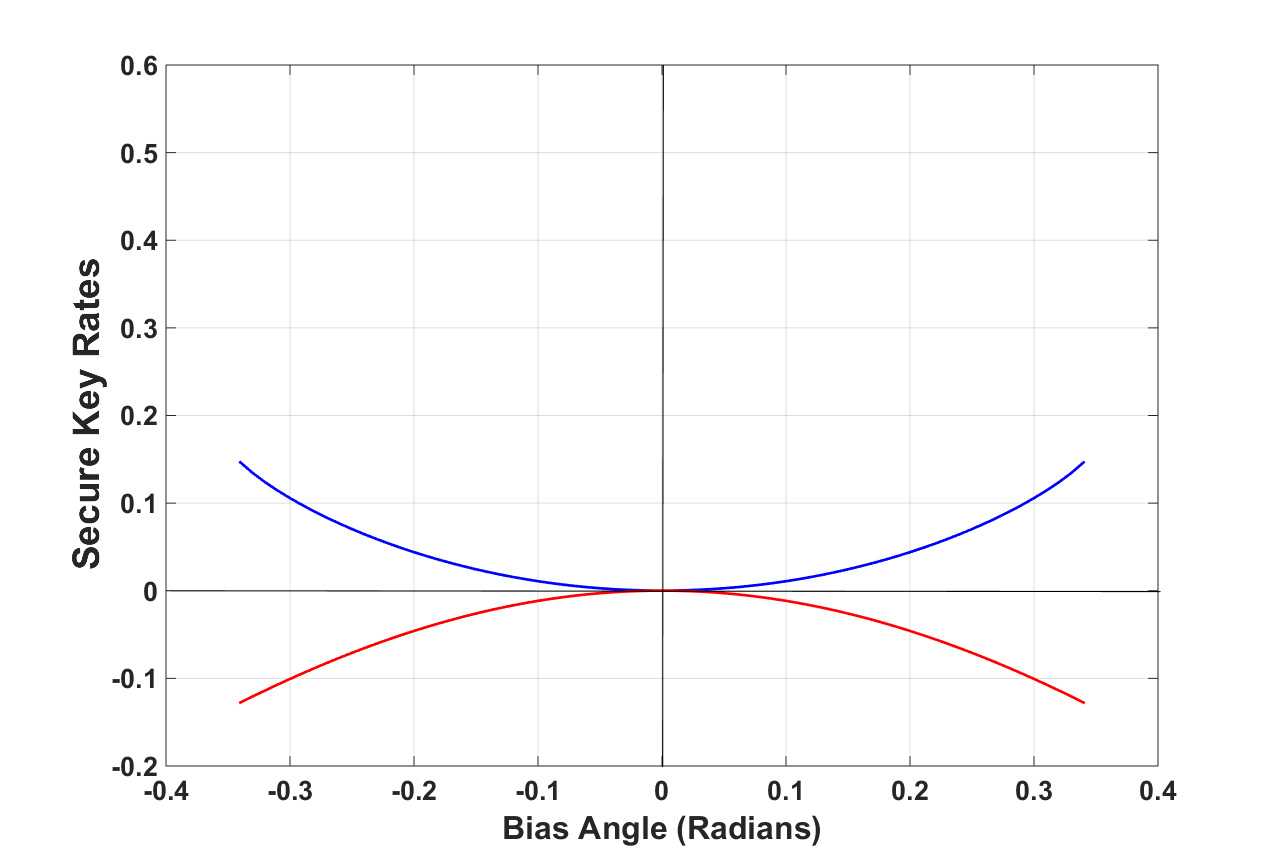}
                \caption{}\label{SubFig:QBER-11}%
        \end{subfigure}%
\caption{The calculated secure key rates $\tilde{R} = 1-H_2(\tilde{\delta}_X)-H_2(\tilde{\delta}_Z)$ (blue) and $\tilde{R} = 1-2H_2(\tilde{\delta}_b)$ (red) as a function of $\varphi$ when the weak measurement observables used to estimate the error rates are given by $\widehat{H}^\pm(\varphi)$. The plots shown correspond to when the signals undergo depolarizing noise with a QBER of (a) $8\%$ and (b) $11\%$. The correct key rate is obtained when $\varphi = 0$. The range of permissible values for $\varphi$ is limited to those for which both $\tilde{\delta}_X$ and $\tilde{\delta}_Z$ are non-negative. As can be seen, by slightly biasing the weak measurement observables in either direction, it is possible to incorrectly determine that the achievable secure key rate is larger than it actually is if we treat the $\delta_X$ and $\delta_Z$ error rates independently. However, by averaging over the two error rate estimates to obtain $\tilde{\delta}_b$, we can overcome this vulnerability.
}\label{Fig:SecKeyRates}%
\end{figure}

Any rotation $\varphi$ that decreases one of the error estimates will always increase the other. We now begin to see the value of using $\delta_b$ as our estimate of the bit error rate, rather than treating $\delta_X$ and $\delta_Z$ separately. To be more precise, if the actual values $\delta_X$ and $\delta_Z$ are nonzero, then there exist a range of values for the rotation $\varphi$ to the weak measurement observables such that the calculated error rates $\tilde{\delta}_X$ and $\tilde{\delta}_Z$ permit the value $\tilde{R} = \max(1-H_2(\tilde{\delta}_X) - H_2(\tilde{\delta}_Z), 0)$ to be strictly greater than the actual asymptotically achievable secure key rate $R = \max(1- H_2(\delta_X) - H_2(\delta_Z), 0)$. See Figure \ref{Fig:SecKeyRates} for some examples of this in the case of depolarizing noise.

If there is a way to independently characterize the rotational bias $\varphi$ applied to the weak measurement observables, then obviously we could infer the correct error rates. Alternatively, the apparent problem can be avoided by averaging the two calculated error rates. In other words, just as in GLLP \cite{GLLP}, we calculate the bit error rate as
\begin{equation}\label{Eq:SmoothQBER}
\tilde{\delta}_b \approx \frac{\tilde{\delta}_X + \tilde{\delta}_Z}{2}.
\end{equation}

This has the effect of ``smoothing out'' weak measurement device imperfections without reducing security. If Eve's attack is basis independent, then we can expect $\delta_X$ and $\delta_Z$ to be approximately equal, so that $\delta_X \approx \delta_Z \approx \delta_b$. Then under this assumption, the calculated asymptotically achievable secure key rate will be given by 
\begin{equation}\label{Eq:SmoothRate}
\tilde{R} = \max(1 - 2H_2(\tilde{\delta}_b), \ 0).
\end{equation}
It turns out that $1 - 2H_2(\tilde{\delta}_b) \leq 1 - H_2(\delta_X) - H_2(\delta_Z)$ for all $\varphi \in (-\pi, \ \pi)$, and in the case of depolarizing noise, equality is reached precisely when $\varphi = 0$, i.e., when there is no biasing. 
%
Thus, we have shown:
\begin{theorem}\label{Thm:BiasedWM}
Security is preserved in the weak measurement QKD protocol in the event that, due either to a malicious Eve or simple device imperfections, the projectors that are actually weakly measured are uniformly biased by an arbitrary angle $\varphi \in (-\pi, \pi)$.
\end{theorem}

We note that we may combine this with the analysis of the previous section. That is to say, it is still possible to obtain a nonzero secure key if the weak measurement observables that are actually measured are from a Gaussian distribution centered around $\wh{H}^\pm(\varphi)$.

An entirely reasonable scenario in which the above biasing can occur is in the event that Alice's source and Bob's detectors are misaligned. In other words, if we suppose that the $\wh{H}$ observables are aligned properly with respect to Bob's detectors, but Alice's source is slightly rotated, then the effect on the weak measurements will be identical to the above scenario.
\subsection{\label{Sec:ChoiceofWMObservable} Adversarial Knowledge of Bob's Choice of Weak Measurement Observable}

Thus far, we have assumed that any attack by Eve on the weak measurement interactions has been independent of Bob's choice of weak measurement observable. Because Bob will randomly choose one of the two observables to weakly measure on each signal in the implementation of the protocol, we will now concern ourselves with the possibility that Eve has some knowledge about this choice and attempts to use it to gain an advantage by selectively and independently rotating the weak measurement observables. We will demonstrate that the protocol is still robust against such weak measurement biasing attacks, \emph{even if Eve has complete knowledge of the choice of observable}. 

Let us suppose that for each signal arriving at Bob's weak measurement interaction, Eve has some probability $p_H \geq \frac{1}{2}$ of correctly guessing Bob's choice of weak measurement observable. With this information, Eve chooses to bias the weak measurement observable one way or another on each signal. With complete information about Bob's choice of weak measurement observables, Eve could potentially rotate $\wh{H}^+$ one way and $\wh{H}^-$ another way in order to try to reduce Alice and Bob's estimated error rates. With only partial information about this choice, Eve must factor in the probability of rotating the wrong observable into her choice of optimal bias.

Let us suppose that Eve tries to rotate $\wh{H}^+$ to $\wh{H}^+(\varphi)$ and $\wh{H}^-$ to $\wh{H}^-(\varphi')$. With probability $p_H \geq \frac{1}{2}$ Eve guesses correctly, in which case the calculated expectation values on each signal become $\la \wh{H}^+(\varphi)\ra_{\rho_\alpha}$ and $\la \wh{H}^-(\varphi')\ra_{\rho_\alpha}$. With probability $(1-p_H)$, Eve guesses the incorrect observable, in which case the calculated expectation values on this subset become $\la \wh{H}^+(\varphi')\ra_{\rho_\alpha}$ and $\la \wh{H}^-(\varphi)\ra_{\rho_\alpha}$.

Eve's goal is to use her knowledge $p_H$ of Bob's choice of weak measurement observable to choose biases $\varphi$ and $\varphi'$ that minimize the estimated error rate $\tilde{\delta}_b$ from Eq. \eqref{Eq:SmoothQBER}. Under the assumption of unital noise, we find that her optimal choice for $\varphi$ and $\varphi'$ is given by 
\begin{align}
\tan(\varphi) &= \frac{r_x^+ - r_z^+(2p_H-1) - r_z^0 + r_x^0(2p_H-1)}{r_x^+ + r_z^+(2p_H-1) + r_z^0 + r_x^0(2p_H-1)},\\
\tan(\varphi') &= \frac{r_x^+ + r_z^+(2p_H-1) - r_z^0 - r_x^0(2p_H-1)}{r_x^+ - r_z^+(2p_H-1) + r_z^0 - r_x^0(2p_H-1)},
\end{align}
where $r_x^\alpha$ and $r_z^\alpha$ are as in Eq. \eqref{Eq:rho-alpha}.

In particular, if the channel is depolarizing, then $r_z^+ = r_x^0 = 0$ and $r_x^+ = r_z^0$, so that the optimal choice of bias becomes $\varphi = \varphi' = 0$. That is to say, if the channel is depolarizing, then any biasing of the weak measurement observables will only serve to increase, rather than decrease the estimated error rate. Furthermore, in the arbitrary unital case, Eve's knowledge $p_H$ of the choice of weak measurement observable is only advantageous to her in the event that there is a rotation applied to the signals, so that $r_z^+$ and $r_x^0$ are nonzero. Let us consider the worst case scenario that Eve has complete knowledge of Bob's choice of weak measurement observable, so that $p_H=1$. We find that, even in this case, under arbitrary unital noise, the estimated error rate $\tilde{\delta}_b$ becomes less than the actual error rate $\delta_b$ only after both error rates are below the $11\%$ security threshold. Consequently, complete knowledge of Bob's choice of weak measurement observable, along with the ability to independently bias each of the weak measurement observables, provides no advantage to Eve. Thus we have shown:
\begin{corollary}
Theorem \ref{Thm:BiasedWM} holds even when Eve has complete knowledge of Bob's choice of weak measurement observable and the ability to independently bias both observables.
\end{corollary}



\subsection{\label{Subsec:ImperfectSources} Weak Coherent Sources}

Since nearly perfect deterministic single photon sources are not yet technologically feasible, a practical implementation of our QKD protocol will use weak coherent laser pulses with randomized phases to approximate single photons. Because even very weak coherent pulses will occasionally contain more than one photon, the weak measurement protocol, as with any other similar prepare and measure protocol, is vulnerable to a Photon Number Splitting (PNS) attack \cite{BLMS2000}. The use of decoy states with different pulse intensities interspersed randomly with the signal pulses was proposed and developed as a method of securing QKD protocols against PNS attacks \cite{LMC2005, Wang2005, MQZL2005,DYDSS2008}. The essential idea is that to successfully extract information from the signal pulses, Eve will need to selectively apply a beamsplitter to pulses with multiple photons in order to to extract information without increasing the bit error rate. This means that a PNS attack will necessarily cause appreciably higher attenuation of pulses with higher intensities \cite{MGCHEMB2015}. This allows Alice and Bob to protect against such an attack by carefully measuring the attenuation of pulses produced with several intensities and using this to place a lower bound on the transmittance of pulses with single photons. When combined with information about the error rate of signal pulses, this allows Alice and Bob to place an upper bound on the error rate of pulses that contain single photons. 

Let us choose a decoy state protocol with a signal pulse of average intensity $\mu$, a decoy state of average intensity $\nu$, with $\nu < \mu$, and a vacuum pulse with zero intensity. By using their knowledge of the pulse intensities and by carefully measuring the probabilities that the signal and decoy pulses produce a detection (i.e., the transmittances), $Q_\mu$ and $Q_\nu$, Alice and Bob can estimate a lower bound for the transmittance for those pulses that contain a single photon, i.e. the ideal case, to be \cite{MQZL2005} 
\begin{equation}
Q_1^L = \frac{\mu^2e^{-\mu}}{\mu\nu - \nu^2}\left[ Q_\nu e^{\nu} - Q_\mu e^{\mu}\left(\frac{\nu}{\mu}\right)^2 -  Q_{vac} \left( \frac{\mu^2-\nu^2}{\mu^2} \right) \right],
\end{equation}
where $Q_{vac}$ is the detector click probability for the vacuum pulses. If we attribute all of the bit errors to those pulses with single photons (the worst case for security since only pulses with single photons contribute to the secure key rate), then an upper bound on the bit error rate of single photon pulses is given by 
\begin{equation}
\varepsilon_1^U = \frac{\varepsilon_\nu Q_\nu e^\nu - \varepsilon_{vac}Q_{vac}}{Q_1^L e^\mu \left( \frac{\nu}{\mu} \right)} \label{Eq:upper_bound_epsilon_single_photon},
\end{equation}
where $\varepsilon_\nu$ and $\varepsilon_{vac}$ are the bit error rates for the decoy and vacuum pulses. From this, using the results of \cite{MQZL2005}, one can calculate that a secure final key can be extracted from the sifted signal pulses at the asymptotic rate given by 
\begin{equation}
R = q \left\{ Q_1^L \left[1 - H_2\left(\varepsilon_1^U\right)\right] - Q_\mu f\left( \varepsilon_\mu \right) H_2\left(\varepsilon_\mu \right) \right\} \label{Eq:key_rate_decoy_usual},
\end{equation}
where $f(\varepsilon_\mu)$ is given by the efficiency of the error reconciliation procedure used in the implementation and $q$ is the fraction of the transmitter signal pulses that are in the sifted key, and $q=1/2$ just as in the standard BB84 protocol since half of the signal pulses will be in the $Z$ basis. Notice that this decoy state analysis implicitly defines the bit error rate to be
\begin{equation}\label{Eq:BER}
\varepsilon = \frac{\delta_X + \delta_Z}{2}=\delta_b.
\end{equation}
All that we need to do in order to arrive at the secure rate of our protocol is estimate the corresponding $\delta_b$ parameters from our weak measurement results for the signal, decoy, and vacuum pulses. In the next section we will show how to adapt the procedure in Section \ref{Sec:PE} to accurately estimate these when the detectors have nonzero dark count rates.

Additionally, if we are somehow certain that the weak measurement observables are implemented with no bias as in Theorem \ref{Thm:PerfectWM}, then we can estimate the corresponding $\delta_X$ and $\delta_Z$ error rates individually. From these we can write an asymptotic bound that is higher than the one above whenever $\delta_X \neq \delta_Z$, given by
\begin{equation}\label{Eq:Rate_WM_with_Decoy}
R = q \left\{ Q_1^L \left[1 - H_2\left(\delta_{X,1}^U\right)\right] - Q_\mu f\left( \delta_{Z,\mu} \right) H_2\left(\delta_{Z,\mu} \right) \right\},
\end{equation}
where $\delta_{Z,\mu}$ is the $Z$ error rate estimated from the weak measurements of the signal pulses and $\delta_{X,1}^U$, the upper bound of the $X$ error rate for pulses containing a single photon, and  is given by
\begin{equation}
\delta_{X,1}^U = \frac{\delta_{X,\nu} Q_\nu e^\nu - \delta_{X,vac} Q_{vac} }{ Q_1^L e^\mu \left( \frac{\nu}{\mu} \right) } \label{Eq:X-error_upper_bound}, 
\end{equation}
where $\delta_{X,\nu}$, and $\delta_{X,vac}$ are the estimated $X$ error rates for the decoy and vacuum pulses respectively. We will now turn to how to accurately estimate the error rates from the weak measurements when the detectors have nonzero dark count rates. 

\subsection{\label{Subsec:ImperfectDetectors} Detector Dark Counts}

Let us suppose that the single-photon detectors in the QKD system have a probability $p_d$ of producing a false detection signal within the detection time window. When there are significant optical losses in the channel, this is the probability that a signal arrived at the detector with zero photons \emph{and} that the detector produced a false positive within the the detection window. For a coherent pulse transmitted with intensity $\gamma$, the probability of it containing zero photons is
\begin{equation}
Q_0^{\gamma} = e^{-\gamma} Y_0 \approx Y_0,
\end{equation}
where $Y_0$ is the yield of the vacuum pulses, i.e. detection events when the vacuum pulses are transmitted. Let $d_i$ be the dark count rate of detector $i$ and $\tau$ be the detection window duration, then $Y_0 = \left(d_1 + d_2\right)\tau$. We can accurately estimate $Y_0$, by measuring the transmittance of the vacuum pulses in the protocol since $Q_{vac} = Y_0 = \left(d_1 + d_2\right)\tau$. For a pulse prepared with an average intensity of $\gamma$ and with an observed transmittance $Q_\gamma$, the fraction of detection events due to detector dark counts, $d(\gamma)$, is given by
\begin{equation}\label{Eq:DarkCountFraction}
d(\gamma) = \frac{\left(d_1 + d_2\right)\tau}{Q_\gamma}.
\end{equation}
We will assume that Eve controls the dark count output of Bob's detectors. Signals arriving at the detectors containing no photons will not contribute to the weak measurements results. In fact, we should expect the average expectation value of the dark count signals for both weak measurement observables to be $0$. Therefore the averages of the weak measurement results for the signal pulses conditioned on the initial state $|\alpha\ra$ and prepared with intensity $\gamma$ become 
\begin{equation}
\mu^\pm_{\alpha,\gamma} = \left[1-d(\gamma) \right] g^\pm\la \wh{H}^\pm \ra_{\rho_\alpha}. \label{Eq:dark_count_WM_avg}
\end{equation}
We see here that the dark counts will effectively make the estimated coupling parameter smaller than its actual value. Since we can estimate the $d(\gamma)$ parameters independently of weak measurement results using our information about the transmitted intensities and the observed transmittances, we can correct for this and accurately extract the expectation values from the observed weak measurements results using Equation \eqref{Eq:dark_count_WM_avg}. 

From the $\mu^\pm_{\alpha,\gamma}$ weak measurement averages, Alice and Bob can estimate the parameters $\tilde{\delta}_{i,\mu}$ and $\tilde{\delta}_{i,\nu}$ with $i \in \{X,Z\}$. However these are the error rates for the fraction of the signals that arrive at the detector with non-zero intensities. To get the total estimated error rate we have to add in the errors due to the observed dark counts with the proper proportion of vacuum and non-vacuum events, 
\begin{equation}\label{Eq:QBER_darkcounts}
\delta_{i,\gamma} = \frac{\tilde{\delta}_{i,\gamma}\left(Q_\gamma - Q_0\right) + \frac{1}{2}Q_0}{Q_\gamma} = \tilde{\delta}_{i,\gamma} + \left(\frac{1}{2} - \tilde{\delta}_{i,\gamma}\right) \frac{\left(d_1 + d_2\right) \tau}{Q_\gamma}, 
\end{equation}
with $\gamma$ the pulse intensity and $i \in \{X,Z\}$. These error rates can then be plugged directly into Equations \eqref{Eq:X-error_upper_bound} and \eqref{Eq:Rate_WM_with_Decoy} or Equations \eqref{Eq:upper_bound_epsilon_single_photon} and \eqref{Eq:key_rate_decoy_usual} to find the asymptotically achievable secure key rate. 

\section{\label{Sec:MDI-security}Towards Weak Measurement Device Independent Security}

Thus far, we have demonstrated the robustness of our protocol against reasonable device flaws and some adversarial control over the weak measurement interactions. While the security of the parameter estimation is independent of Bob's post-selection, and therefore detector control by Eve, there remains the possibility that Eve could compromise the protocol's parameter estimation if she could somehow control the weak measurement outcomes themselves. In this section we will demonstrate that, given a very weak restriction on Eve's information about which observable is weakly measured on each signal, the weak measurement protocol can withstand a very general and powerful class of attack strategies in which Eve has complete control of the weak measurement outcomes. 


In addition to Assumption \ref{Asm:PathManipulation}, which prevents Eve from performing projective measurements of each photon \emph{after} the weak measurement interaction and before the photon detectors, we will impose the following additional requirement on Eve's knowledge regarding the weak measurements:
\begin{assumption}\label{Asm:H-Uncertainty}
Eve has limited information about which observable, $\wh{H}^+$ or $\wh{H}^-$, Bob has chosen to weakly measure on each signal. More precisely, we assume that Eve's ability to correctly guess which of the two observables Bob chooses to measure on each signal is bounded above by some probability $\frac{1}{2} \leq p_H < 1$.
\end{assumption}
The precise limitation of Eve's knowledge required will be derived below.

In order to model Eve's control over the weak measurement results, we will treat the weak measurement device in the QKD system as a black box through which each signal passes immediately prior to being strongly measured at Bob's detectors. We will imagine that the inner working of the black box can be controlled by an agent of Eve, who we will called Fred. In addition to controlling the weak measurement outcomes, Fred is also assumed to have full information about Bob's choice of which observable he intends to be weakly measured for each signal. Eve is allowed to communicate freely with Fred, but Fred will have some limitation imposed on his ability to communicate with Eve. This limitation will impose a limit on Eve's probability, $p_H$, of correctly guessing Bob's weak measurement observable choice for each signal. 

The general weak measurement attack strategy we will consider is the following: (1) Eve will perform any attack on the quantum channel she desires, (2) Fred attempts to communicate the weak measurement observable choice to Eve, then (3) Eve uses her available information to instruct Fred on how to influence the weak measurement results that Bob records for each signal. Whether or not these results are due to an actual measurement interaction is unimportant. Eve's goal is to influence the weak measurement results such that her interactions with the quantum channel are hidden, and therefore Alice and Bob will certify the channel as secure. 


The weak measurement protocol is not fully measurement-device independent; nevertheless, the black box model of the weak measurement outcomes given above is somewhat analogous to the channel error rate estimation procedure in MDI-QKD. In MDI-QKD, both Alice and Bob's signals are handed to an adversary, Charlie, who is responsible for making a Bell state measurement. Charlie has the option of either publicly announcing the actual measurement result, or announcing some other value of his choosing. In order to verify that the results Charlie announces are correlated with the transmitted signals, Alice and Bob authenticate the measurement results by comparing to the expected results of the Hong-Ou-Mandel (HOM) interference of the two signal photons. Charlie's lack of knowledge about both Alice and Bob's signals prevent him from being able to fake the measurement results and pass the HOM interference test.

In the WM-QKD protocol, there is no HOM interference test. However, we conjecture that given the validity of Assumptions \ref{Asm:PathManipulation} and \ref{Asm:H-Uncertainty}, Alice and Bob can still verify that the announced weak measurement results are correlated with the transmitted signals by placing three basic restrictions on the measurement results that can be easily verified as additional steps in the protocol. 
%
%
The first restriction is that the estimates for $\delta_X$ and $\delta_Z$ are both nonnegative. In the nonadversarial regime, this requirement will always be satisfied; however, this imposes a rather strong restriction on Eve's ability to fake the weak measurement results when she has limited knowledge of Bob's choice of weak measurement observable. The second check is that the variance of the measurement results is independent of the source. That is to say, the calculated variances of the weak measurement results conditioned on each initial state should be statistically identical. As we will see, Eve's attempts to fake the weak measurement results can sometimes affect the variance of the measurement outcomes differently depending on the source basis. The last check is that the variance of the weak measurement results are what we expect given the prescribed weak measure pointer state and interaction. That is to say, we use Eq. \eqref{Eq:WM-Var} to bound the permissible variance of the weak measurement results. As we will show, an adversary cannot simultaneously fake the weak measurement results and satisfy all of these checks, even with limited access to both source basis information and Bob's choice of weak measurement observable. We will call these checks the ``weak measurement verification'' or ``WM verification.'' Our goal in the following is to demonstrate that, with limited side channel access to the source basis and Bob's choice of weak measurement observable, any weak measurement attack strategy by Eve will either fail the WM verification or generate error rate estimates above the security threshold.

Before characterizing several general attack strategies in detail, let us first write down the expectation values of $\wh{H}^+$ and $\wh{H}^-$ for each of the four input states with no channel noise. When the signal is encoded with bit value $a$ in the $Z$ basis and there is no channel noise, the expectation values are:
\begin{equation}\label{Eq:EveWMZ}
\la \wh{H}^\pm \ra_{a} = \frac{1}{2} + \frac{\left(-1\right)^a}{2\sqrt{2}}.
\end{equation}
While when the signal is encoded in the $X$ basis with no channel noise, we have:
\begin{equation}\label{Eq:EveWMX}
\la \wh{H}^\pm \ra_{a} = \frac{1}{2} \pm \frac{\left(-1\right)^a}{2\sqrt{2}}.
\end{equation}

As can be seen, the expectation values (and hence the weak measurement results) have encoded in them information about the source bit value. Consequently, one may be led to believe that the extent to which Eve can fake these expectation values will be limited by her knowledge of the source \emph{bit} values. However, Eve can just as easily make a weak measurement of both observables on each signal \emph{before} she performs any additional interaction with the signal, and while Eve does not know the bit values, she does know that her weak measurement results are correlated with the raw key bit values. Consequently, we come to the first of three distinct attack strategies that utilize Eve's control of the weak measurement device: 

\medskip

\noindent\textbf{Attack Strategy 1} ($p_{basis}=1/2$, $p_H\leq 1$). 
If Eve knows with certainty the weak measurement observable chosen for each signal, then Eve should make her own weak measurements of both $\widehat{H}^+$ and $\widehat{H}^-$ near the signal source and record these results. After her weak measurements of the signals, Eve then performs a strong $Z$ basis measurement on the quantum channel and records the results. Finally, Eve uses her control of the weak measurement device to substitute her weak measurement results corresponding to the observable that Bob expects, while sending a copy of her strong measurement result to Bob's detectors. 

This attack strategy is a very limited one since it requires $p_H=1$. A natural question is whether or not Eve can utilize her own weak measurement results and her control over Bob's weak measurement device to compromise the protocol when $1/2<p_H<1$. Let us assume for now that Eve has no knowledge of Alice's preparation basis, so that her probability of correctly guessing the basis is $p_{basis} = 1/2$. We conjecture that the most general strategy for this scenario is the following one. Eve replaces Bob's weak measurement results with a scalar multiple (actually an affine transformation) of her weak measurement results, in an attempt to leverage her knowledge of Bob's choice of weak measurement observable to manipulate the error estimates without any \emph{a priori} knowledge of the encoded bit values.

Note two crucial observations regarding the expectation values from Eqs. \eqref{Eq:EveWMZ} and \eqref{Eq:EveWMX}. First, when the signals are encoded in the $Z$ basis the expectation values of $\wh{H}^+$ and $\wh{H}^-$ are the same, while those encoded in the $X$ basis have a sign difference in the second term. Second, by assumption, Eve does not have any information about the basis in which a given signal was prepared. 

In the case of unital noise, we remind the reader that the $X$ and $Z$ errors are related to the true expectation values by
\begin{align}
\delta_X &= \frac{1-r_x^+}{2} = \frac{1+r_x^-}{2},\\
\delta_Z &= \frac{1-r_z^0}{2} = \frac{1+r_z^1}{2},
\end{align}
where the $r$ parameters are obtained from
\begin{align}
r_x^+ &= \sqrt{2}(\la H^+\ra_+ - \la H^-\ra_+)\\ 
r_z^0 &= \sqrt{2}(\la H^+\ra_0 + \la H^-\ra_0 - 1).
\end{align}

Eve has the goal of altering Alice and Bob's estimate of these parameters such that the average of the two is as small as possible (or at least below the security threshold). As before, we will denote Alice and Bob's estimates of this as
\begin{equation}
\tilde{\delta}_b = \frac{\tilde{\delta}_X + \tilde{\delta}_Z}{2} = \frac{1}{2} - \frac{1}{4} \left(\tilde{r}_x^+ + \tilde{r}_z^0 \right),
\end{equation}
where the tilde over the $r$'s indicate these are derived from the fake weak measurement outcomes provided by Eve. Note that in the case of no channel noise, 
\begin{align} 
r_x^+ &= +1, \ r_x^- = -1,\\ 
r_z^0 &= +1, \ r_z^1 = -1.
\end{align}
This fact will form the basis of Eve's attack strategy to leverage her weak measurement results to attempt to break the protocol. If Eve only correctly guesses Bob's choice of weak measurement observable with probability $p_H<1$ and she knows nothing about the source basis, she can choose the weak measurement results given to Bob so that
\begin{equation}
\Delta_\text{pointer}^\text{Fake}(H^\pm) = \frac{g_{Eve}}{2} + \alpha \left(\Delta_\pm^{Eve} - \frac{g_{Eve}}{2}\right).
\end{equation}
In the above, $g_{Eve}$ is the coupling parameter for Eve's weak measurement interactions. We will assume that Eve is smart enough to set $g_{Eve} \approx g$ in order not to raise the suspicions of Alice and Bob. From these fake weak measurement pointer results, Alice and Bob's estimates of the $r$ parameters are given by
\begin{align}
\tilde{r}_x^+ &= \sqrt{2}(\la H^+\ra_+^\text{Fake} - \la H^-\ra_+^\text{Fake}),\\ 
\tilde{r}_z^0 &= \sqrt{2}(\la H^+\ra_0^\text{Fake} + \la H^-\ra_0^\text{Fake} - 1),
\end{align}
which Eve wishes to minimize. The faked weak measurement results will yield faked estimated expectation values given by 
\begin{align} 
\la \wh{H}^+ \ra_{a}^\text{Fake} &= \frac{1}{2} + \alpha \left[ p_H \left(\la \wh{H}^+ \ra_{a} - \frac{1}{2} \right) + (1-p_H) \left( \la \wh{H}^- \ra_{a} - \frac{1}{2} \right) \right] \label{Eq:FakeWM+} \\
\la \wh{H}^- \ra_{a}^\text{Fake} &= \frac{1}{2} + \alpha \left[ p_H \left(\la \wh{H}^- \ra_{a} - \frac{1}{2} \right) + (1-p_H) \left( \la \wh{H}^+ \ra_{a} - \frac{1}{2} \right) \right] \label{Eq:FakeWM-},
\end{align}
where $\la \wh{H}^{\pm} \ra_{a}$ are from the results of Eve's weak measurements and given in Equations \eqref{Eq:EveWMZ} and \eqref{Eq:EveWMX}. From this Alice and Bob's $r$ parameter estimates will now be 
\begin{align}
\tilde{r}_x^+ &= +\alpha \left( 2p_H - 1 \right) \\
\tilde{r}_x^- &= -\alpha \left( 2p_H - 1 \right) \\
\tilde{r}_z^0 &= + \alpha \\
\tilde{r}_z^1 &= - \alpha.
\end{align}
Therefore the bit error rate estimated by Alice and Bob under this weak measurement attack will be 
\begin{equation}\label{Eq:FakeErrorEst-Attack1}
\tilde{\delta}_b = \frac{1}{2} \left( 1 - \alpha p_H \right).
\end{equation}
This looks good for Eve since $\alpha$ is freely chosen by her. However, Alice and Bob will test the trustworthiness of the weak measurement device by estimating the variance of the weak measurement results. Recognizing that the weak measurements are independent of the source, the estimated variances conditioned on each source should be statistically equivalent.

The variances of the fake measurement results conditioned on the source basis are given by 
\begin{align}
\label{Eq:FakeVarX}
\text{Var}\left[\Delta_\text{pointer}^\text{Fake}(H^{\pm},X)\right] &= \text{Var}\left[\alpha \Delta_{\pm,x}^\text{Eve}\right] = \left[ \alpha\frac{g_{Eve} \sigma_{Eve}}{\left(2p_H - 1\right)} \right]^2 \\
\label{Eq:FakeVarZ}
\text{Var}\left[\Delta_\text{pointer}^\text{Fake}(H^{\pm},Z)\right] &= \text{Var}\left[\alpha \Delta_{\pm,z}^\text{Eve}\right] = \left( \alpha g_{Eve} \sigma_{Eve} \right)^2. 
\end{align}
The variance test performed by Alice and Bob then forces Eve to make Eqs. \eqref{Eq:FakeVarX} and \eqref{Eq:FakeVarZ} equal, which can only be satisfied when $p_H=1$. We note that this is the degenerate case covered by Attack Strategy 1. The significance of this result is that if Eve has no information about the source basis, then in the asymptotic key regime only perfect information about the weak measurement observable will allow her to utilize her control of the weak measurement device to compromise the security of the protocol.

\medskip

\noindent\textbf{Attack Strategy 2} ($\frac{1}{2}<p_{basis}<1$, $\frac{1}{2}<p_H<1$).
With the previous result we are lead to ask to what extent Eve can use any side-channel information about the source basis \emph{combined} with imperfect information about the weak measurement observable to break the protocol. It turns out that such a strategy does exist, and the details of this attack and the derivation of the resulting lower bound on Alice and Bob's estimated QBER can be found in Appendix \ref{App:GeneralAttack}. This lower bound is given by

\begin{equation}\label{Eq:LowerBound_Attack2}
\tilde{\delta}_b^L = \frac{1}{2} \left[1 -  \left(\frac{\sigma_{sec}}{\sigma_{MD}}\right) p_{basis}p_H \right].
\end{equation}

Given an error security bound of $\delta_{sec}$, the protocol is secure against this attack strategy whenever $\delta_{sec} - \delta_{wm} < \tilde{\delta}_b^L$. In other words, the ratio $\frac{\sigma_{sec}}{\sigma_{MD}}$ will place an upper bound on the allowed value of the product $p_{basis}p_H$ such that the protocol's security is preserved when this attack is employed. 

\medskip

\noindent\textbf{Attack Strategy 3} ($p_{basis}> 1-2\delta_{sec}$, $p_H<1$).
In this case, Eve's most powerful attack strategy is to simply perform an intercept-resend attack on the quantum channel, measuring in the $Z$ basis whenever she believes the signal was prepared in $Z$, and leave Bob's true weak measurement results unaltered. Since $p_{basis} > 1-2\delta_{sec}$, it follows that the noise introduced by guessing the incorrect basis will be below the error security bound. Obviously, this attack is one open to Eve against any QKD protocol that is not completely device independent. 


We therefore make the following conjecture: 

\begin{conjecture}\label{Conj:OptimalAttack}
In the limit of infinite key length and given that Assumptions \ref{Asm:PathManipulation} and \ref{Asm:H-Uncertainty} are valid, either Attack Strategy 2 or Attack Strategy 3 is optimal for all values of $p_{basis}$ and $p_H$. In other words, given the assumptions, there are no attack strategies that utilize source basis and weak measurement observable side-channel information that extract a greater amount of key information with an equivalent QBER as estimated by Alice and Bob. 
\end{conjecture}

The most likely method Eve would use to gain information about the weak measurement observable choice is via an active Trojan horse attack on Bob's weak measurement apparatus. These attacks would be implemented using essentially the same techniques as Trojan horse attacks on the source or measurement basis choices \cite{Jain14} in other QKD protocols. Characterizing and limiting information leakage into such side channels is a requirement for any QKD protocol that is not fully device independent. In particular, we note that the MDI-QKD protocol is also vulnerable to Trojan horse attacks on either Alice or Bob's sources. In effect, MDI-QKD trades the vulnerability to detector attacks on Bob's detection basis choice in BB84 for vulnerability to Trojan horse attacks on Bob's source basis choice. Our protocol utilizes a similar trade by swapping Bob's detector basis choice for his choice of which observable to weakly measure. A key point is that our protocol is able to make this trade using only single photons and without any need for successful two photon interference as in MDI-QKD. In addition, Eve needs a significantly larger amount of information about the choice of observable in WM-QKD to break the protocol than she needs about Bob's choice of source basis in MDI-QKD. The full protocol including the weak measurement certification step is included in Appendix \ref{App:Protocol}.

\section{\label{Sec:Implementation}Implementation}

To indicate the feasibility of the protocol we will now sketch one particular implementation.  Let us suppose that the raw key information is encoded in the polarization of the signal photons with Alice randomly choosing to encoded in either the linear polarization basis ($Z$) or the diagonal basis ($X$). Furthermore, the weak measurement pointer state will be the temporal wavefunction of the signal pulses. The weak measurement pointer states are prepared so that the single photons are emitted with a temporal envelope that is Gaussian shaped with a fixed width $\tau$ and a peak value that occurs at a known time as determined by an accurate clock at the source. Bob will possess a clock that is synchronized with Alice's source clock. Bob uses this clock to record the time each photon is detected by the single photon detectors that perform the final $Z$ basis measurement. In the new protocol Bob must also weakly measure one of the projectors onto the two $H$ polarization states (randomly chosen) given by
\begin{eqnarray}
\vert H^\pm \rangle = \cos\left(\pm \frac{\pi}{8}\right)|0\rangle - \sin\left(\pm \frac{\pi}{8}\right)|1\rangle.
\end{eqnarray}

\begin{figure}
\centering
 \includegraphics[width=0.75\linewidth]{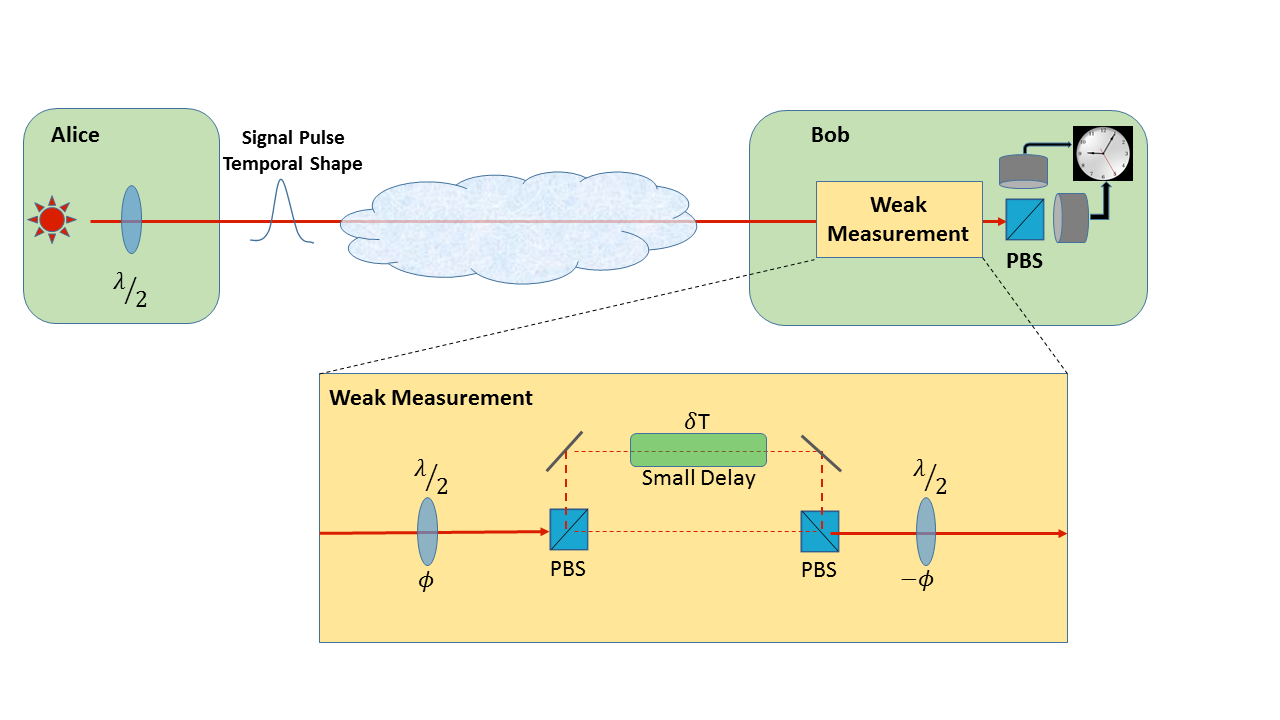}
\caption{Schematic of one possible implementation of the weak measurement QKD protocol. Key information is encoded using the polarization of the signal photons. The photons' time degree of freedom is used as the weak measurement pointer. The weak measurement coupling strength is controlled by the size of the relative delay between the paths in the interferometer. 
}\label{Fig:WM_Imp}%
\end{figure}

To implement the necessary weak measurement interaction, Bob will use a polarizing interferometer consisting of two polarizing beam splitters oriented with the linear basis and a polarization independent optical delay in one path of the interferometer with duration $\delta T \ll \tau$ as in Figure \ref{Fig:WM_Imp}. Because the optical delay is small compared with the photon's temporal uncertainty, the entanglement between the photon's polarization and path is very weak. Just before each photon enters the polarizing interferometer, Bob will uniformly randomly choose to measure one of the two projectors onto $\widehat{H}^+$ and $\widehat{H}^-$ by rotating the incoming photon's polarization state by $\phi=\frac{\pi}{8}$ or $\phi=-\frac{\pi}{8}$ respectively. On exiting the interferometer, the polarization of each photon is then rotated back to its original state.  Note that by implementing the weak measurement interaction in this way, the interaction strength is ensured to always be the same for both projectors since the same delay is used. The weak measurement pointer used here is the time degree of freedom for the photons' exiting of the source aperture. As long as the signal pulse's temporal wavefunction is well controlled and the detectors' timing jitter are relatively small compared to the weak measurement's optical delay, accurate estimates of delay time, and thus the weak measurement results, can be obtained. 

Finally, we note that this implementation is only one of many possibilities. For example, in the case that the raw key information is encoded into the phase of the signal pulses, the polarization degree of freedom could be used as the weak measurement pointer. Bob would then utilize weak coupling between the paths inside of the appropriate non-polarizing interferometer and the photons' polarization to implement a weak measurement of phase. Additionally, while we have chosen to highlight using an additional degree of freedom of the photons themselves for the weak measurement pointer states, in principle these states could be the states of some other quantum systems prepared by Bob. Bob would weakly couple this system to each photon as it is received and then strongly measure this system independently of the photons. One potential method of using an external quantum system to perform the weak measurements is to to perform a suitably weak non-demolition measurement of the photon number of the path through the interferometer defining the weak measurement observable. This could potentially be accomplished using a nonlinear optical interaction to shift the phase of a coherent state local oscillator used as the weak measurement pointer \cite{Gaeta12,Sinclair16} in the presence of a signal photon in the interferometer path. A nice feature of this application of a nonlinear optical interaction based non-demolition measurement is that, in contrast to performing high quality quantum gates for quantum computing, we do not need or even desire a high degree of distinguisability between the occupation and non-occupation of the path. 

A benefit of the analysis of the previous section is that by implementing the weak measurements using additional degrees of freedom of the photons themselves, and thus allowing Eve direct access to the pointer states, we do not necessarily compromise the security of the protocol. 

\section{Performance Comparisons of Predicted Asymptotic Secure Key Rate}\label{Sec:Performance}

\begin{figure}
\centering
 \includegraphics[width=0.75\linewidth]{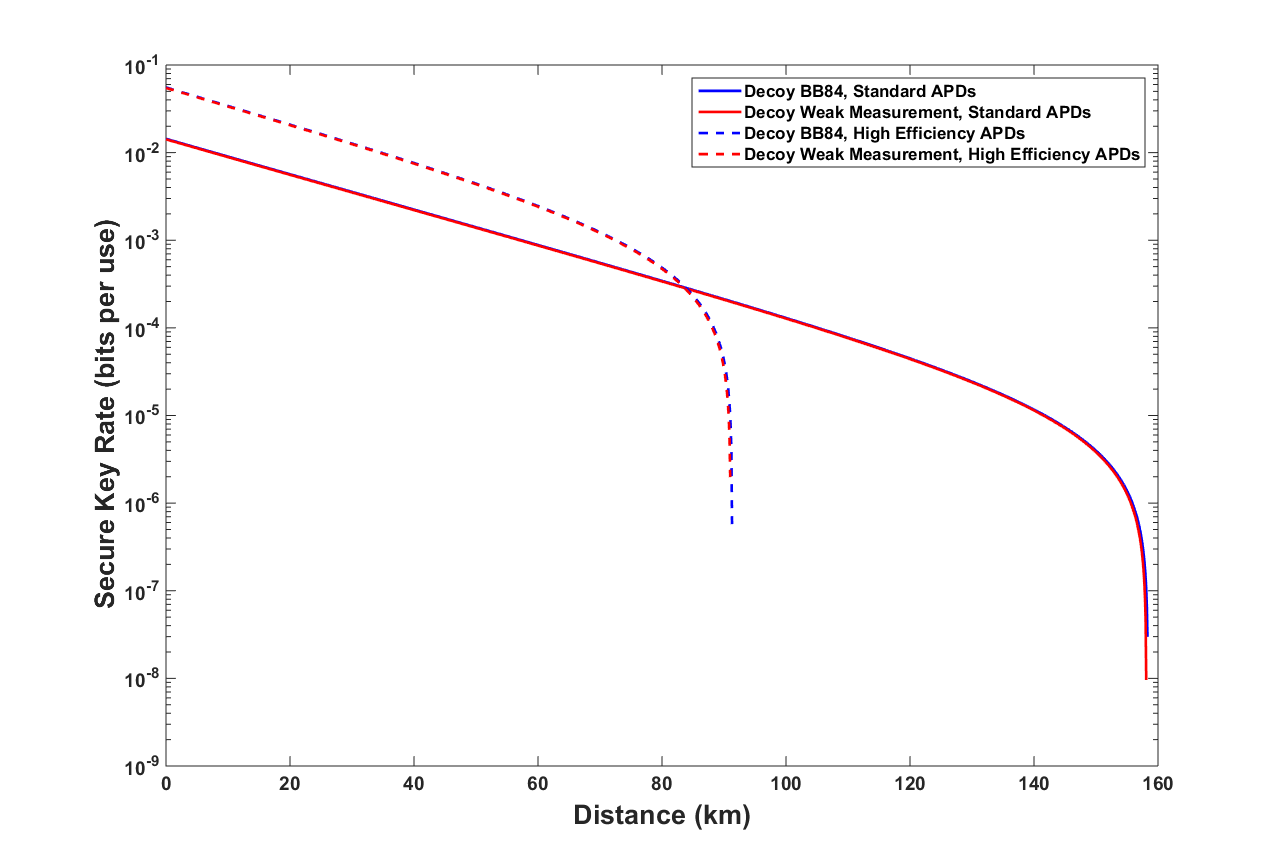}
\caption{Comparison of secure key rates of Weak Measurement and BB84 with protocols, both with Weak+Vacuum decoy states. The pulse intensities for each protocol are $\mu=0.48$, $\nu=0.05$, channel loss rate is $0.2$ dB/km, and error reconciliation factor is $f=1.22$. Solid lines use measurement devices with with total detection efficiency $\eta_d=0.145$ and vacuum count rate $Y_0=6\times10^{-6}$. For dotted lines, $\eta_d=0.55$ and $Y_0=5\times10^{-4}$. The weak measurement strength is $\sigma/g=0.05$. The system intrinsic error rate in both cases is 1.5\%.
}\label{Fig:WM-vs-BB84}%
\end{figure}

In this section we will directly compare the asymptotic secure key rates of the new weak measurement protocol, a decoy state BB84 protocol, and the MDI protocol of Lo, \emph{et al.} under various QKD system configurations. In Figure \ref{Fig:WM-vs-BB84} we compare the Decoy+Vacuum BB84 protocol, calculated using Equation (42) in \cite{MQZL2005}, and the Weak+Vacuum Weak Measurement protocol, calculated using Equation (\ref{Eq:Rate_WM_with_Decoy}), with two types of avalanche photon detectors (APDs): standard InGaAs APDs \cite{Gobby04} and higher efficiency InGaAs/InP APDs \cite{Comandar2015}. We have chosen realistically weak interactions for the weak measurements with $g/\sigma_{MD}=0.05$, as well as a conservative intrinsic system error rate of $e_d=0.015$, which we model as a rotation of the state of the source qubit in the $X,Z$ plane as defined by the weak measurment observables. The detection efficiency and dark count rate parameters were taken from the QKD experiment reported in \cite{Gobby04}. Notice that the penalty we pay in secure rate when using weak measurements for parameter estimation is barely perceptible. This shows that the weak measurement protocol has almost the same asymptotic secure key rate as BB84 itself, but without vulnerabilities to detector control by Eve. 

\begin{figure}
\centering
    \begin{subfigure}[b]{0.5\textwidth}
    		\centering
                \includegraphics[width=0.95\linewidth]{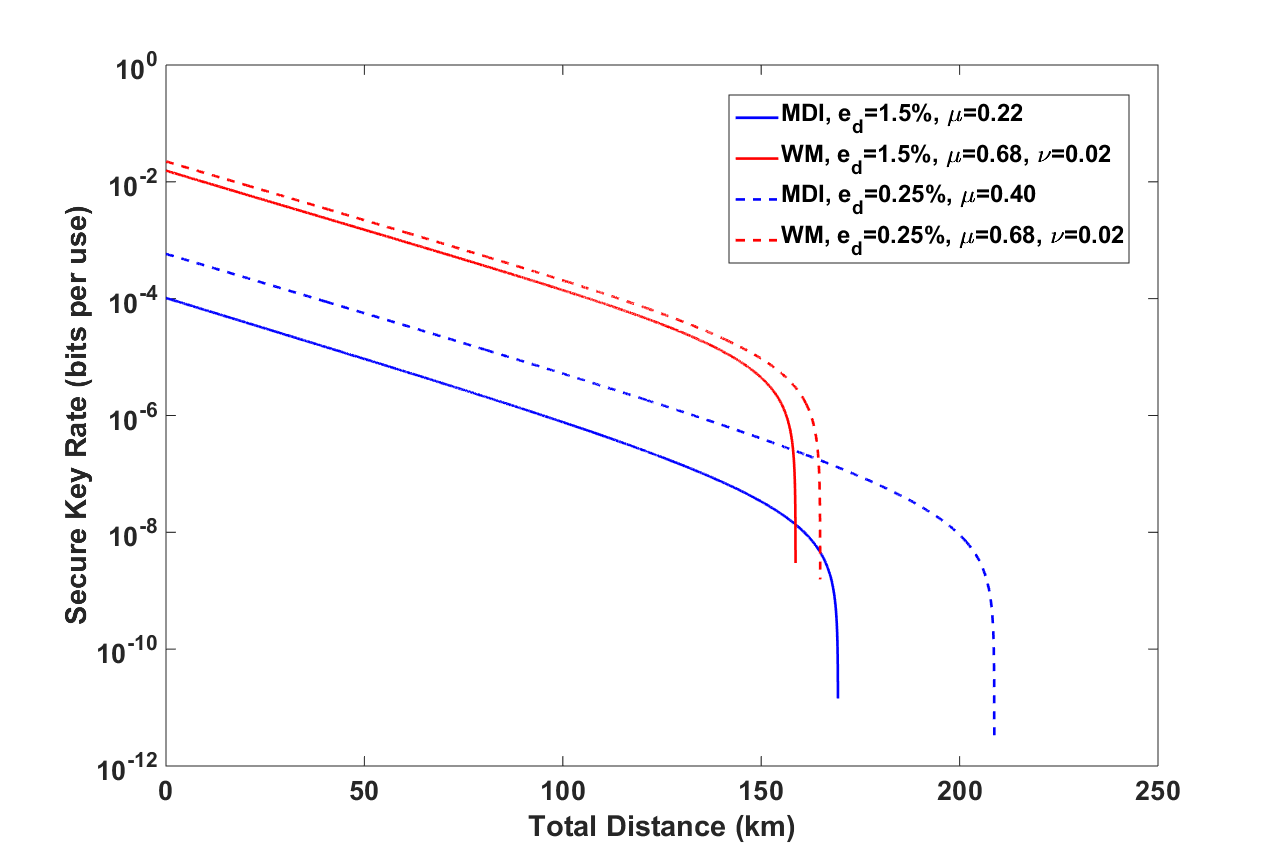}
                \caption{}\label{SubFig:SecKey_Detectors_WM}%
        \end{subfigure}%
    \begin{subfigure}[b]{0.5\textwidth}
        	\centering
                \includegraphics[width=0.95\linewidth]{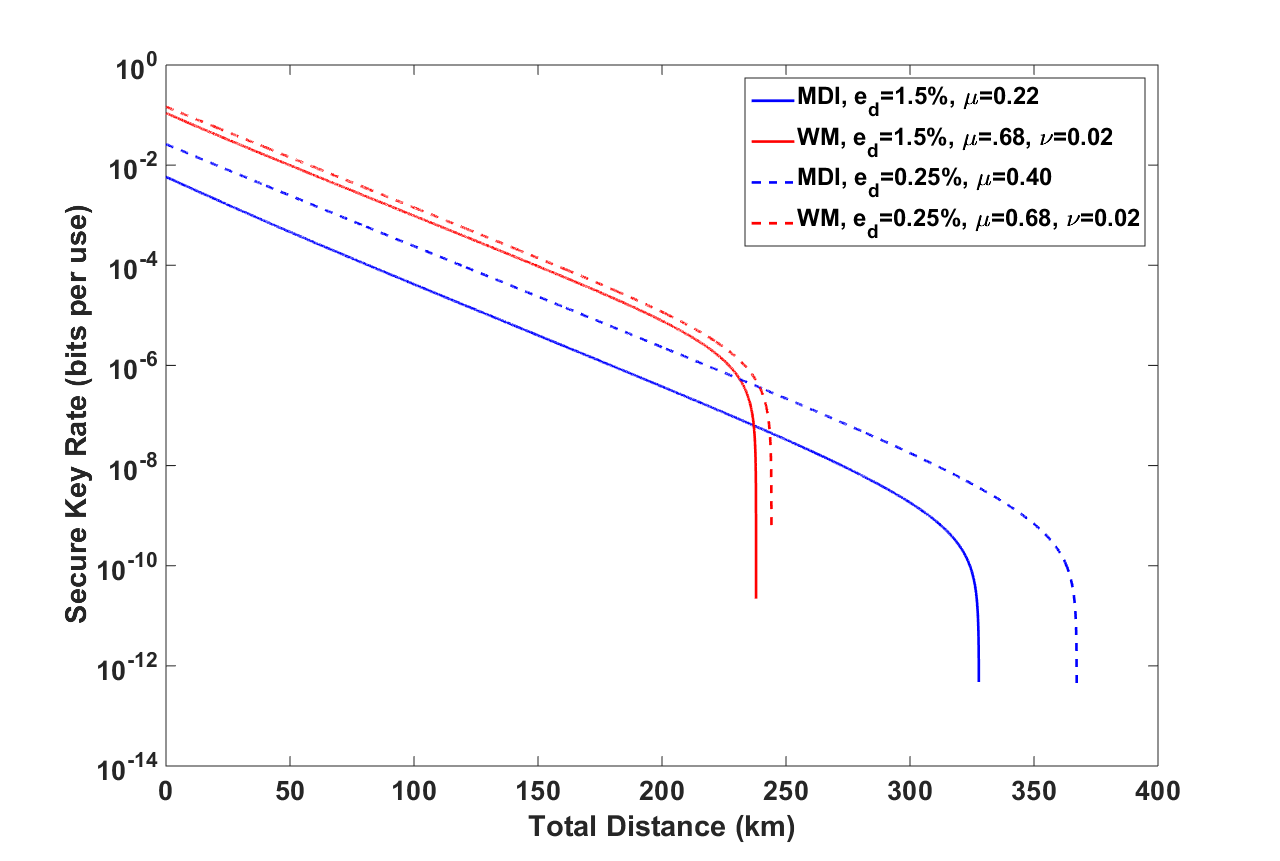}
                \caption{}\label{SubFig:SecKey_MDIvWM_SNSPD}%
        \end{subfigure}%
\caption{Comparison of WM-QKD and MDI-QKD secure key rates using (a) standard, non-cryogenic APDs with total detection efficiency of $14.5\%$ \cite{Gobby04} and (b) cryogenic SNSPDs with total detection efficiency of $93\%$ \cite{Marsili13} for moderate (solid curve) and very low (dotted curve) intrinsic system error rates. The channel loss rate is $0.2$ dB/km and the error reconciliation factor is $f=1.16$.
}\label{Fig:SecKey_MDI-WM}%
\end{figure}

Since WM-QKD is secure against all detector control attacks, it is instructive to compare its performance to MDI-QKD, currently the most feasible way to implement quantum cryptography without any detector vulnerabilities. In the limit of an infinite number of decoy states and assuming the only intrinsic system errors are due to rotations of the source qubits, the asymptotic secure key rate of MDI-QKD can be calculated as in \cite{Xu15}. In Figure \ref{Fig:SecKey_MDI-WM} we compare the relative performance of the two protocols for very low and moderate intrinsic system errors of $0.25\%$ and $1.5\%$ respectively. Using standard APDs, Figure \ref{SubFig:SecKey_Detectors_WM}, indicates that WM-QKD has a secure key rate that is typically two orders of magnitude higher than MDI-QKD and a positive secure rate for a distance that is about 75-90\% that of MDI-QKD. In Figure \ref{SubFig:SecKey_MDIvWM_SNSPD} we compare WM-QKD and MDI-QKD when using cryogenic superconducting nanowire single-photon detectors with 93\% detection efficiency \cite{Marsili13}. As expected, the use of such extremely high efficiency detectors significantly increases the performance of both protocols, and also closes the gap somewhat between the two in the case of very low intrinsic system error. In all of these cases we have attempted to optimize the choice of signal and/or decoy intensities to maximize the key rates for each protocol and each set of system parameters.  

\begin{figure}
\centering
 \includegraphics[width=0.8\linewidth]{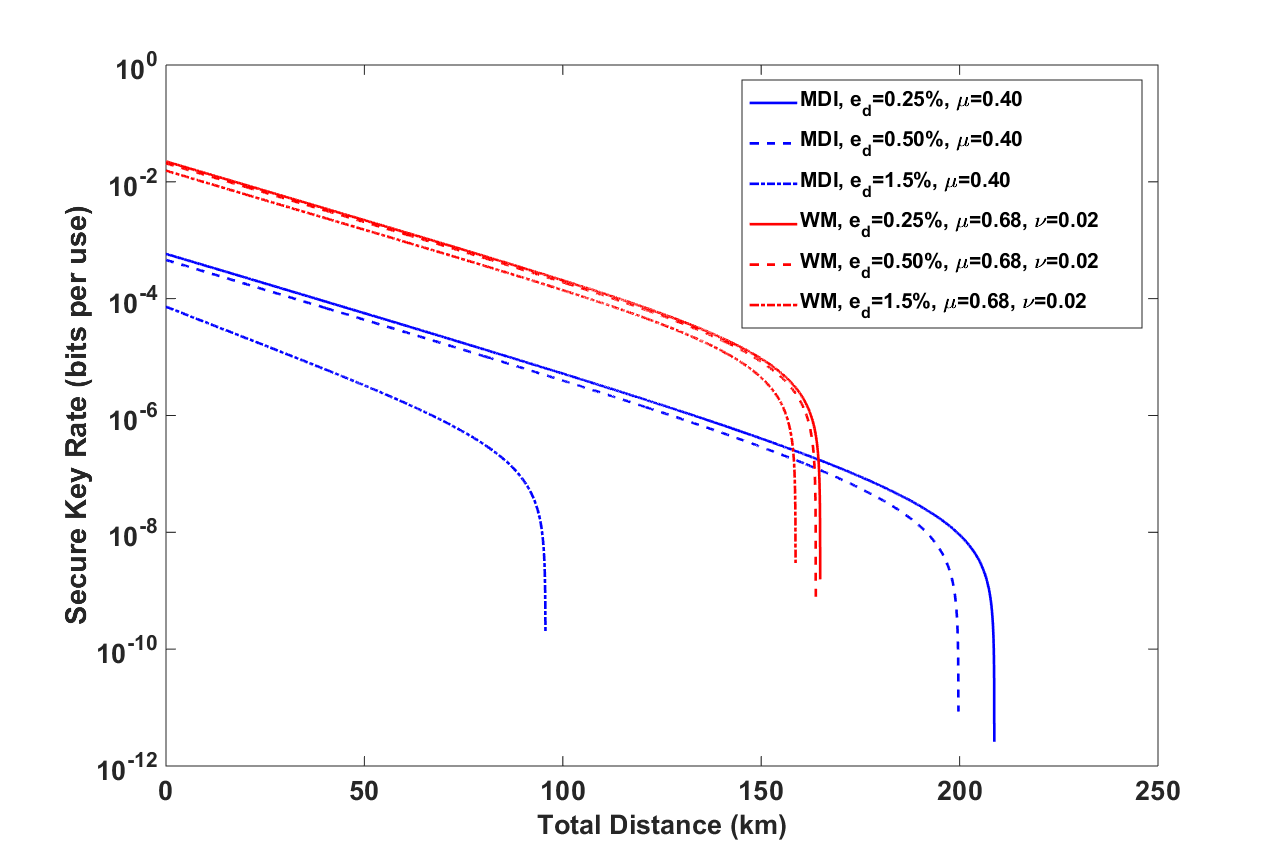}
\caption{This figure shows the stability of WM-QKD as compared to MDI-QKD under increases in the intrinsic system error rate. The total detection efficiency is $14.5\%$, the channel loss rate is $0.2$ dB/km, and the error reconciliation factor is $f=1.16$.
}\label{Fig:Stability_WMvMDI}%
\end{figure}

An important issue for the performance of a practical QKD system is the stability of the secure key rate with respect to changes in the system parameters. Due to the complex relationship between these parameters in MDI-QKD there can be a significant variability in the achievable secure key rate with small changes in source intensity or system error rate. For example, in Figure \ref{Fig:Stability_WMvMDI} we compare how changes in the intrinsic system error rate affect the secure key rate for MDI-QKD and WM-QKD. The signal/decoy intensities for each protocol were optimized for a system intrinsic error rate of $0.25\%$ and held fixed. As the system's error rate was increased above $1\%$, MDI-QKD's performance degrades very rapidly. In comparison, the performance of WM-QKD is quite stable as a function of the intrinsic system error. This is due to the relative simplicity of performing single qubit measurements compared to the Bell state measurements of MDI-QKD.        

\section{Conclusion}

We have presented a new prepare-and-measure QKD protocol that utilizes weak measurements to decouple security parameter estimation from the photon detection basis. This protocol provides a much simpler solution to well-known detector control attacks than alternatives such as MDI-QKD. We have demonstrated the robustness of the protocol against weak measurement implementation imperfections. 
Additionally, we have analyzed the extreme scenario in which Eve can completely control the weak measurement results themselves, and conjectured optimal attack strategies in this regime. We have demonstrated that these attack strategies necessarily fail when there is a small limitation on Eve's ability to accurately guess Bob's choice of weak measurement observable and Eve cannot alter the photon's state \emph{after} the the weak measurement interaction. Finally, we compared the asymptotically achievable secure key rates of the new weak measurement protocol to BB84 with Weak+Vacuum decoy states and MDI-QKD with an infinite number of decoy states. We found that the weak measurement protocol should have almost identical performance to the BB84 protocol and is up to two orders of magnitude better than MDI-QKD for realistic system parameters and using standard, non-cryogenic detectors. 

While MDI-QKD protocols eliminate detectors from the analysis of security, implementation of these protocols are highly non-trivial, costly, and have a reduced overall key throughput as compared to prepare-and-measure protocols. The prepare-and-measure weak measurement QKD protocol introduced in this article offers a cost-effective alternative to MDI-QKD that, while also eliminating detector control based attacks, will achieve an overall secure key throughput essentially equivalent to the original BB84 protocol. We stress that although the weak measurement protocol is not fully detection device independent because it is in principle susceptible to some detector side-channels, e.g. timing attacks \cite{Lamas07}, that can leak key information independent of channel parameter estimation, WM-QKD \emph{is} immune to detector control, e.g. detector blinding attacks. 

The analysis presented in this paper proves the asymptotic security of a weak measurement QKD protocol (1) in the ideal case, (2) with reasonable device imperfections, and (3) against a general class of attacks in which Eve has complete control over the outcomes of the weak measurements. Future work will address the security of the weak measurement QKD protocol in the finite key length regime. In addition, we will work to develop detailed descriptions of realistic implementations of the protocol and derive the resulting predicted secure key rates. 

\section*{Acknowledgments}

J. Troupe acknowledges support from the Office of Naval Research Code 31, grant number N00014-15-1-2225. JT also thanks the Institute for Quantum Studies at Chapman University and the Perimeter Institute for Theoretical Physics for their hospitality as a portion of this work was completed during visits to each.  J. Farinholt acknowledges support from the Office of Naval Research Code 31, grant number N0001416WX01474, and an NSWCDD In-house Laboratory Independent Research (ILIR) grant.
\bibliographystyle{plainnat}
\bibliography{bibfile_20170109}

\begin{thebibliography}{40}
\providecommand{\natexlab}[1]{#1}
\providecommand{\url}[1]{\texttt{#1}}
\expandafter\ifx\csname urlstyle\endcsname\relax
  \providecommand{\doi}[1]{doi: #1}\else
  \providecommand{\doi}{doi: \begingroup \urlstyle{rm}\Url}\fi

\bibitem[Aharonov and Vaidman(1990)]{AV90}
Yakir Aharonov and Lev Vaidman.
\newblock Properties of a quantum system during the time interval between two
  measurements.
\newblock \emph{Physical Review A}, 41:\penalty0 11--20, Jan 1990.
\newblock \doi{10.1103/PhysRevA.41.11}.
\newblock URL \url{http://link.aps.org/doi/10.1103/PhysRevA.41.11}.

\bibitem[Aharonov et~al.(1988)Aharonov, Albert, and Vaidman]{AAV1988}
Yakir Aharonov, David~Z. Albert, and Lev Vaidman.
\newblock How the result of a measurement of a component of the spin of a spin-
  \textit{1/2} particle can turn out to be 100.
\newblock \emph{Phys. Rev. Lett.}, 60:\penalty0 1351--1354, Apr 1988.
\newblock \doi{10.1103/PhysRevLett.60.1351}.
\newblock URL \url{http://link.aps.org/doi/10.1103/PhysRevLett.60.1351}.

\bibitem[Barrett et~al.(2005)Barrett, Hardy, and Kent]{Barrett05}
Jonathan Barrett, Lucien Hardy, and Adrian Kent.
\newblock No signaling and quantum key distribution.
\newblock \emph{Physical Review Letters}, 95:\penalty0 010503, Jul 2005.
\newblock \doi{10.1103/PhysRevLett.95.010503}.

\bibitem[Barrett et~al.(2013)Barrett, Colbeck, and Kent]{Barrett13}
Jonathan Barrett, Roger Colbeck, and Adrian Kent.
\newblock Memory attacks on device-independent quantum cryptography.
\newblock \emph{Physical Review Letters}, 110:\penalty0 010503, Jan 2013.
\newblock \doi{10.1103/PhysRevLett.110.010503}.

\bibitem[Belavkin and Staszewski(1992)]{BS92}
V.~P. Belavkin and P.~Staszewski.
\newblock Nondemolition observation of a free quantum particle.
\newblock \emph{Physical Review A}, 45:\penalty0 1347--1356, Feb 1992.
\newblock \doi{10.1103/PhysRevA.45.1347}.
\newblock URL \url{http://link.aps.org/doi/10.1103/PhysRevA.45.1347}.

\bibitem[Bennett and Brassard(1984)]{BB84}
Charles Bennett and Gilles Brassard.
\newblock Quantum cryptography: Public key distribution and coin tossing.
\newblock In \emph{Proceedings of the IEEE International Conference on
  Computers, Systems, and Signal Processing}, page 175, 1984.

\bibitem[Brassard et~al.(2000)Brassard, L\"utkenhaus, Mor, and
  Sanders]{BLMS2000}
Gilles Brassard, Norbert L\"utkenhaus, Tal Mor, and Barry~C. Sanders.
\newblock Limitations on practical quantum cryptography.
\newblock \emph{Physical Review Letters}, 85:\penalty0 1330--1333, Aug 2000.
\newblock \doi{10.1103/PhysRevLett.85.1330}.
\newblock URL \url{http://link.aps.org/doi/10.1103/PhysRevLett.85.1330}.

\bibitem[Carmichael et~al.(1991)Carmichael, Brecha, and Rice]{CBR91}
H.J. Carmichael, R.J. Brecha, and P.R. Rice.
\newblock Quantum interference and collapse of the wavefunction in cavity qed.
\newblock \emph{Optics Communications}, 82\penalty0 (1):\penalty0 73 -- 79,
  1991.
\newblock ISSN 0030-4018.
\newblock \doi{http://dx.doi.org/10.1016/0030-4018(91)90194-I}.
\newblock URL
  \url{http://www.sciencedirect.com/science/article/pii/003040189190194I}.

\bibitem[Carmichael()]{Car93}
Howard Carmichael.
\newblock \emph{An open systems approach to quantum optics}, volume~18.
\newblock Springer-Verlag Berlin Heidelberg.
\newblock \doi{10.1007/978-3-540-47620-7}.

\bibitem[Comandar et~al.(2015)Comandar, Fr{\"o}lich, Dynes, Lucamarini, Yuan,
  Penty, and Shields]{Comandar2015}
L.C. Comandar, B.~Fr{\"o}lich, J.F. Dynes, M.~Lucamarini, Z.L. Yuan, R.V.
  Penty, and A.J. Shields.
\newblock Gigahertz-gated {I}n{G}a{A}s/{I}n{P} single-photon detector with
  efficiency exceeding 55
\newblock \emph{Journal of Applied Physics}, 117:\penalty0 083109, 2015.
\newblock \doi{10.1063/1.4913527}.

\bibitem[Cortez et~al.(2017)Cortez, Chantasri, Garc\'{\i}a-Pintos, Dressel, and
  Jordan]{CCetal17}
Luis Cortez, Areeya Chantasri, Luis~Pedro Garc\'{\i}a-Pintos, Justin Dressel,
  and Andrew~N. Jordan.
\newblock Rapid estimation of drifting parameters in continuously measured
  quantum systems.
\newblock \emph{Physical Review A}, 95:\penalty0 012314, Jan 2017.
\newblock \doi{10.1103/PhysRevA.95.012314}.
\newblock URL \url{http://link.aps.org/doi/10.1103/PhysRevA.95.012314}.

\bibitem[Curty et~al.(2014)Curty, Xu, Cui, Lin, Tamaki, and Lo]{Curty14}
Marcos Curty, Feihu Xu, Wei Cui, Charles~Wen Lin, Kiyoshi Tamaki, and Hoi-Kwong
  Lo.
\newblock Finite-key analysis for measurement-device-independent quantum key
  distribution.
\newblock \emph{Nature Communications}, 5:\penalty0 3732, 2014.
\newblock \doi{10.1038/ncomms4732}.

\bibitem[Dixon et~al.(2008)Dixon, Yuan, Dynes, Sharpe, and Shields]{DYDSS2008}
A.~R. Dixon, Z.~L. Yuan, J.~F. Dynes, A.~W. Sharpe, and A.~J. Shields.
\newblock Gigahertz decoy quantum key distribution with 1 mbit/s secure key
  rate.
\newblock \emph{Opt. Express}, 16\penalty0 (23):\penalty0 18790--18797, Nov
  2008.
\newblock \doi{10.1364/OE.16.018790}.
\newblock URL
  \url{http://www.opticsexpress.org/abstract.cfm?URI=oe-16-23-18790}.

\bibitem[Gerhardt et~al.(2011)Gerhardt, Liu, Lamas-Linares, Skaar, Scarani,
  Makarov, and Kurtsiefer]{Gerh11}
Ilja Gerhardt, Qin Liu, Ant\'{i}a Lamas-Linares, Johannes Skaar, Valerio
  Scarani, Vadim Makarov, and Christian Kurtsiefer.
\newblock Experimentally faking the violation of bell's inequalities.
\newblock \emph{Physical Review Letters}, 107:\penalty0 170404, Oct 2011.
\newblock \doi{10.1103/PhysRevLett.107.170404}.
\newblock URL \url{http://link.aps.org/doi/10.1103/PhysRevLett.107.170404}.

\bibitem[Gobby et~al.(2004)Gobby, Yuan, and Shields]{Gobby04}
C.~Gobby, Z.L. Yuan, and A.J. Shields.
\newblock Quantum key distribution over 122 km of standard optical fiber.
\newblock \emph{Applied Physics Letters}, 84:\penalty0 3762, May 2004.
\newblock \doi{10.1063/1.1738173}.

\bibitem[Gonz\'{a}z et~al.(2015)Gonz\'{a}z, Reb\'{o}n, da~Silva, Saavedra,
  Curty, Lima, Xavier, and Nogueira]{Gonzalez15}
P.~Gonz\'{a}z, L.~Reb\'{o}n, T.~Ferreira da~Silva, M.~Saavedra, M.~Curty,
  G.~Lima, G.B. Xavier, and W.A.T. Nogueira.
\newblock Quantum key distribution with untrusted detectors.
\newblock \emph{Physical Review A}, 92:\penalty0 022337, 2015.
\newblock \doi{10.1103/PhysRevA.92.022337}.

\bibitem[Gottesman et~al.(2004)Gottesman, Lo, L\"{u}tkenhaus, and
  Preskill]{GLLP}
Daniel Gottesman, Hoi-Kwong Lo, Norbert L\"{u}tkenhaus, and John Preskill.
\newblock Security of quantum key distribution with imperfect devices.
\newblock \emph{Quantum Information and Computation}, 4\penalty0 (5):\penalty0
  325--360, Sept 2004.
\newblock ISSN 1533-7146.
\newblock URL \url{http://dl.acm.org/citation.cfm?id=2011586.2011587}.

\bibitem[Jain et~al.(2014)Jain, Anisimova, Khan, Makarov, Marquardt, and
  Leuchs]{Jain14}
Nitin Jain, Elena Anisimova, Imran Khan, Vadim Makarov, Christoph Marquardt,
  and Gerd Leuchs.
\newblock Trojan-horse attacks threaten the security of practical quantum
  cryptography.
\newblock \emph{New Journal of Physics}, 16:\penalty0 123030, Oct 2014.
\newblock \doi{10.1088/1367-2630/16/12/123030}.

\bibitem[Lamas-Linares and Kurtsiefer(2007)]{Lamas07}
Ant\'{i}a Lamas-Linares and Christian Kurtsiefer.
\newblock Breaking a quantum key distribution system through a timing side
  channel.
\newblock \emph{Optics Express}, 15:\penalty0 9388--9393, Jul 2007.
\newblock \doi{10.1364/OE.15.009388}.

\bibitem[Lo and Chau(1999)]{LoChau}
Hoi-Kwong Lo and H.~F. Chau.
\newblock Unconditional security of quantum key distribution over arbitrarily
  long distances.
\newblock \emph{Science}, 283\penalty0 (5410):\penalty0 2050--2056, 1999.
\newblock ISSN 0036-8075.
\newblock \doi{10.1126/science.283.5410.2050}.
\newblock URL \url{http://science.sciencemag.org/content/283/5410/2050}.

\bibitem[Lo et~al.(2005)Lo, Ma, and Chen]{LMC2005}
Hoi-Kwong Lo, Xiongfeng Ma, and Kai Chen.
\newblock Decoy state quantum key distribution.
\newblock \emph{Physical Review Letters}, 94:\penalty0 230504, Jun 2005.
\newblock \doi{10.1103/PhysRevLett.94.230504}.
\newblock URL \url{http://link.aps.org/doi/10.1103/PhysRevLett.94.230504}.

\bibitem[Lo et~al.(2012)Lo, Curty, and Qi]{LoCurQi12}
Hoi-Kwong Lo, Marcos Curty, and Bing Qi.
\newblock Measurement-device-independent quantum key distribution.
\newblock \emph{Physical Review Letters}, 108:\penalty0 130503, Mar 2012.
\newblock \doi{10.1103/PhysRevLett.108.130503}.
\newblock URL \url{http://link.aps.org/doi/10.1103/PhysRevLett.108.130503}.

\bibitem[Lydersen et~al.(2010)Lydersen, Wiechers, Wittmann, Elser, Skaar, and
  Makarov]{Lyd10}
Lars Lydersen, Carlos Wiechers, Christoffer Wittmann, Dominique Elser, Johannes
  Skaar, and Vadim Makarov.
\newblock Hacking commercial quantum cryptography systems by tailored bright
  illumination.
\newblock \emph{Nature Photonics}, 4:\penalty0 686 -- 689, 2010.
\newblock \doi{10.1038/nphoton.2010.214}.
\newblock URL \url{http://dx.doi.org/10.1038/nphoton.2010.214}.

\bibitem[Ma et~al.(2005)Ma, Qi, Zhao, and Lo]{MQZL2005}
Xiongfeng Ma, Bing Qi, Yi~Zhao, and Hoi-Kwong Lo.
\newblock Practical decoy state for quantum key distribution.
\newblock \emph{Physical Review A}, 72:\penalty0 012326, Jul 2005.
\newblock \doi{10.1103/PhysRevA.72.012326}.
\newblock URL \url{http://link.aps.org/doi/10.1103/PhysRevA.72.012326}.

\bibitem[Mailloux et~al.(2015)Mailloux, Grimaila, Colombi, Hodson, Engle,
  McLaughlin, and Baumgartner]{MGCHEMB2015}
L.~O. Mailloux, M.~R. Grimaila, J.~M. Colombi, D.~D. Hodson, R.~D. Engle, C.~V.
  McLaughlin, and G.~Baumgartner.
\newblock Quantum key distribution: examination of the decoy state protocol.
\newblock \emph{IEEE Communications Magazine}, 53\penalty0 (10):\penalty0
  24--31, Oct 2015.
\newblock ISSN 0163-6804.
\newblock \doi{10.1109/MCOM.2015.7295459}.

\bibitem[Makarov et~al.(2016)Makarov, Bourgoin, Chaiwongkhot, Gagn\'{e},
  Jennewein, Kaiser, Kashyap, Legr\'{e}, Minshull, and Sajeed]{Makarov16}
Vadim Makarov, Jean-Philippe Bourgoin, Poompong Chaiwongkhot, Mathiey
  Gagn\'{e}, Thomas Jennewein, Sarah Kaiser, Raman Kashyap, Mathieu Legr\'{e},
  Carter Minshull, and Shihan Sajeed.
\newblock Creation of backdoors in quantum communications via laser damage.
\newblock \emph{Physical Review A}, 94:\penalty0 030302(R), 2016.
\newblock \doi{https://doi.org/10.1103/PhysRevA.94.030302}.

\bibitem[Marsili et~al.(2013)Marsili, Verma, Stern, Harrington, Lita, Gerrits,
  Vayshenker, Baek, Shaw, Mirin, and Nam]{Marsili13}
F.~Marsili, V.B. Verma, J.A. Stern, S.~Harrington, A.E. Lita, T.~Gerrits,
  I.~Vayshenker, B.~Baek, M.D. Shaw, R.P. Mirin, and S.W. Nam.
\newblock Detecting single infrared photons with 93
\newblock \emph{Nature Photonics}, 7:\penalty0 210--214, Mar 2013.
\newblock \doi{10.1038/nphoton.2013.13}.

\bibitem[Sajeed et~al.(2016)Sajeed, Huang, Sun, Xu, Makarov, and
  Curty]{Sajeed16}
Shihan Sajeed, Anqi Huang, Shihai Sun, Feihu Xu, Vadim Makarov, and Marcos
  Curty.
\newblock Insecurity of detector-device-independent quantum key distribution.
\newblock \emph{Physical Review Letters}, 117:\penalty0 250505, 2016.
\newblock \doi{https://doi.org/10.1103/PhysRevLett.117.250505}.

\bibitem[Shor and Preskill(2000)]{ShorPreskill}
Peter~W. Shor and John Preskill.
\newblock Simple proof of security of the {{BB}}84 quantum key distribution
  protocol.
\newblock \emph{Physical Review Letters}, 85:\penalty0 441--444, Jul 2000.
\newblock \doi{10.1103/PhysRevLett.85.441}.
\newblock URL \url{http://link.aps.org/doi/10.1103/PhysRevLett.85.441}.

\bibitem[Silva et~al.(2015)Silva, Gisin, Guryanova, and Popescu]{Silva15}
Ralph Silva, Nicolas Gisin, Yelena Guryanova, and Sandu Popescu.
\newblock Multiple observers can share the nonlocality of half of an entangled
  pair by using optimal weak measurements.
\newblock \emph{Physical Review Letters}, 114:\penalty0 250401, Jun 2015.
\newblock \doi{10.1103/PhysRevLett.114.250401}.
\newblock URL \url{http://link.aps.org/doi/10.1103/PhysRevLett.114.250401}.

\bibitem[Sinclair et~al.(2016)Sinclair, Heshami, Oblak, Simon, and
  Tittel]{Sinclair16}
N.~Sinclair, K.~Heshami, D.~Oblak, C.~Simon, and W.~Tittel.
\newblock Proposal and proof-of-principle demonstration of non-destructive
  detection of photonic qubits using a {T}m:{L}i{N}b$\text{{O}}_3$ waveguide.
\newblock \emph{Nature Communications}, 7:\penalty0 13454, Nov 2016.
\newblock \doi{10.1038/ncomms13454}.

\bibitem[Tang et~al.(2014)Tang, Yin, Chen, Liu, Zhang, Jiang, Zhang, Wang, You,
  Guan, Yang, Wang, Liang, Zhang, Zhou, Ma, Chen, Zhang, and Pan]{Tang14}
Yan-Lin Tang, Hua-Lei Yin, Si-Jing Chen, Yang Liu, Wei-Jun Zhang, Xiao Jiang,
  Lu~Zhang, Jian Wang, Li-Xing You, Jian-Yu Guan, Dong-Xu Yang, Zhen Wang, Hao
  Liang, Zhen Zhang, Nan Zhou, Xiongfeng Ma, Teng-Yun Chen, Qiang Zhang, and
  Jian-Wei Pan.
\newblock Measurement-device-independent quantum key distribution over 200 km.
\newblock \emph{Physical Review Letters}, 113:\penalty0 190501, Nov 2014.
\newblock \doi{10.1103/PhysRevLett.113.190501}.
\newblock URL \url{http://link.aps.org/doi/10.1103/PhysRevLett.113.190501}.

\bibitem[Valivarthi et~al.(2015)Valivarthi, Lucio-Martinez, Chan, Rubenek,
  John, Korchinski, Duffin, Marsili, Verma, Shaw, Stern, Nam, Oblak, Zhou,
  Slater, and Tittel]{Tittel15}
Raju Valivarthi, Itzel Lucio-Martinez, Philip Chan, Allison Rubenek, Caleb
  John, Daniel Korchinski, Cooper Duffin, Francesco Marsili, Varun Verma,
  Mathew Shaw, Jeffrey Stern, Sae~Woo Nam, Daniel Oblak, Qiang Zhou, Joshua
  Slater, and Wolfgang Tittel.
\newblock Measurement-device-independent quantum key distribution: from idea
  towards application.
\newblock \emph{Journal of Modern Optics}, 62:\penalty0 1141--1150, 2015.
\newblock \doi{10.1080/09500340.2015.1021725}.

\bibitem[Venkataraman et~al.(2012)Venkataraman, Saha, and Gaeta]{Gaeta12}
Vivek Venkataraman, Kasturi Saha, and Alexander Gaeta.
\newblock Phase modulation at the few-photon level for weak-nonlinearity-based
  quantum computing.
\newblock \emph{Nature Photonics}, 7:\penalty0 138--141, Feb 2012.
\newblock \doi{10.1038/nphoton.2012.283}.

\bibitem[von Neumann(1955)]{vonNeumann55}
John von Neumann.
\newblock \emph{Mathematical {{F}}oundations of {{Q}}uantum {{M}}echanics}.
\newblock Princeton University Press, 1955.

\bibitem[Wang(2005)]{Wang2005}
Xiang-Bin Wang.
\newblock Beating the photon-number-splitting attack in practical quantum
  cryptography.
\newblock \emph{Physical Review Letters}, 94:\penalty0 230503, Jun 2005.
\newblock \doi{10.1103/PhysRevLett.94.230503}.
\newblock URL \url{http://link.aps.org/doi/10.1103/PhysRevLett.94.230503}.

\bibitem[Weber et~al.(2014)Weber, Chantasri, Dressel, Jordan, Murch, and
  Siddiqi]{WCDJMS14}
S.~J. Weber, A.~Chantasri, J.~Dressel, A.~N. Jordan, K.~W. Murch, and
  I.~Siddiqi.
\newblock Mapping the optimal route between two quantum states.
\newblock \emph{Nature}, 511:\penalty0 570 -- 573, July 2014.
\newblock \doi{10.1038/nature13559}.

\bibitem[Wiseman(2002)]{Wiseman02}
H.~M. Wiseman.
\newblock Weak values, quantum trajectories, and the cavity-{QED} experiment on
  wave-particle correlation.
\newblock \emph{Physical Review A}, 65:\penalty0 032111, Feb 2002.
\newblock \doi{10.1103/PhysRevA.65.032111}.
\newblock URL \url{http://link.aps.org/doi/10.1103/PhysRevA.65.032111}.

\bibitem[Xu et~al.(2013)Xu, Curty, Qi, and Lo]{Bing13}
Feihu Xu, Marcos Curty, Bing Qi, and Hoi-Kwong Lo.
\newblock Practical aspects of measurement-device-independent key distribution.
\newblock \emph{New Journal of Physics}, 15:\penalty0 113007, 2013.
\newblock \doi{10.1088/1367-2630/15/11/113007}.

\bibitem[Xu et~al.(2015)Xu, Curty, Qi, Qian, and Lo]{Xu15}
Feihu Xu, Marcos Curty, Bing Qi, Li~Qian, and Hoi-Kwong Lo.
\newblock Discrete and continuous variables for measurement-device-independent
  quantum cryptography (and {S}upplementary {I}nformation).
\newblock \emph{Nature Photonics}, Correspondence, Nov 2015.
\newblock \doi{10.1038/nphoton.2015.206}.

\end{thebibliography}


\appendix

\section{Analysis of Weak Measurement Attack Strategy 2\label{App:GeneralAttack}}

In this appendix we perform a detailed analysis of Attack Strategy 2 in Section \ref{Sec:MDI-security}. The idea is to generalize Attack Strategy 1 by utilizing Eve's information about the source basis to enable Eve to apply two independent affine transformations to her weak measurement results. With this extra degree of freedom to exploit in designing the faked weak measurements, Eve will attempt to reduce Alice and Bob's estimated QBER below the security threshold and simultaneously satisfy the constraints of the expected variance of the weak measurement results. 

Let the probability that Eve correctly guesses the encoding basis for a given signal be $p_\text{basis} > \frac{1}{2}$. Then, the averages of Eve's weak measurements conditioned on the basis labels that Eve applies to her weak measurement results -- and labeled by the correct $\wh{H}$ observable expected by Bob will be given by 
\begin{align}\label{Eq:WMxAvg}
\Delta_{a,x}^{\pm Eve} &= g_{Eve} \la \wh{H}^\pm \ra_{a,x}^{Eve} = g_{Eve}\left[ \frac{1}{2} \pm p_{basis} \frac{(-1)^a}{2\sqrt{2}} + (1-p_{basis})\frac{(-1)^a}{2\sqrt{2}} \right],  \\ \label{Eq:WMzAvg}
\Delta_{a,z}^{\pm Eve} &= g_{Eve} \la \wh{H}^\pm \ra_{a,z}^{Eve} = g_{Eve}\left[ \frac{1}{2} + p_{basis} \frac{(-1)^a}{2\sqrt{2}} \pm (1-p_{basis})\frac{(-1)^a}{2\sqrt{2}}  \right].
\end{align}
However for an individual signal, Eve is not certain of which of the two $\wh{H}$ observables she needs to fake the weak measurement. By Assumption \ref{Asm:H-Uncertainty}, Eve can only correctly guess this with some probability, $\frac{1}{2}<p_H<1$. Given her knowledge, Eve will guess the observable Bob expects to see the weak measurement of and the preparation basis of the signal, and then she will provide a fake weak measurement result to Bob based on her own weak measurement pointer results for $\wh{H}^+$ and $\wh{H}^-$, denoted $\Delta_+^{Eve}$ and $\Delta_-^{Eve}$, respectively. For a given signal, let us suppose Eve believes it to have been prepared in the $X$ basis and that Bob expects a weak measurement result for $\wh{H}^\pm$. Then Eve will provide a fake weak measurement pointer result to Bob using her own weak measurement pointer result given by
\begin{equation}
\Delta_\text{pointer}^\text{Fake}(H^\pm,X) = \frac{g_{Eve}}{2} + \alpha_x \left(\Delta_\pm^{Eve} - \frac{g_{Eve}}{2}\right),
\end{equation}
for some $\alpha_x$ of her choosing. Likewise, if Eve believes the signal was prepared in the $Z$ basis, she will give Bob the fake weak measurement pointer result 
\begin{equation}
\Delta_\text{pointer}^\text{Fake}(H^\pm,Z) = \frac{g_{Eve} }{2} + \alpha_z \left(\Delta_\pm^{Eve} - \frac{g_{Eve}}{2}\right),
\end{equation}
where again, $\alpha_z$ is some number of Eve's choosing. In the above, $g_{Eve}$ is the coupling parameter for Eve's weak measurement interactions. We will assume that Eve is smart enough to set $g_{Eve} \approx g$ in order not to raise the suspicions of Alice and Bob.

After being provided these fake weak measurement results, the averages calculated by Alice and Bob for a given observable and prepared in a given basis will be
\begin{align}
\la \wh{H}^\pm \ra_{a,x}^\text{Fake} &= \frac{1}{2} + \alpha_x \Big\{ p_{basis} \left[p_H(\pm 1) + \left(1-p_H\right)(\mp 1) \right] + \left(1-p_{basis}\right) \Big\} \frac{\left(-1\right)^a}{2\sqrt{2}} \\
 &= \frac{1}{2} + \alpha_x \Big\{ \left(1-p_{basis} \right) \pm \left(2p_H - 1\right)p_{basis} \Big\} \frac{(-1)^a}{2\sqrt{2}}
\end{align}
and
\begin{align}
\la \wh{H}^\pm \ra_{a,z}^\text{Fake} &= \frac{1}{2} + \alpha_z \Big\{ p_{basis} + \left(1-p_{basis}\right)\left[p_H(\pm 1) + \left(1-p_H\right)(\mp 1) \right] \Big\} \frac{\left(-1\right)^a}{2\sqrt{2}} \\
 &= \frac{1}{2} + \alpha_z \Big\{ p_{basis} \pm \left(1-p_{basis}\right)\left(2p_H - 1\right) \Big\} \frac{\left(-1\right)^a}{2\sqrt{2}}.
\end{align}
This will yield four equations to determine how Eve chooses the faked weak measurement results labelled by the observable Eve believes Bob is expecting and the basis in which she believes Alice prepared the signal:
\begin{align}
\la \wh{H}^+ \ra_{a,x}^\text{Fake} &= \frac{1}{2} + \alpha_x \left[ p \left(\la \wh{H}^+ \ra_{a,x} - \frac{1}{2} \right) + (1-p) \left( \la \wh{H}^- \ra_{a,x} - \frac{1}{2} \right) \right] \\
\la \wh{H}^- \ra_{a,x}^\text{Fake} &= \frac{1}{2} + \alpha_x \left[ p \left(\la \wh{H}^- \ra_{a,x} - \frac{1}{2} \right) + (1-p) \left( \la \wh{H}^+ \ra_{a,x} - \frac{1}{2} \right) \right] \\
\la \wh{H}^+ \ra_{a,z}^\text{Fake} &= \frac{1}{2} + \alpha_z \left[ p \left(\la \wh{H}^+ \ra_{a,z} - \frac{1}{2} \right) + (1-p) \left( \la \wh{H}^- \ra_{a,z} - \frac{1}{2} \right) \right] \\
\la \wh{H}^- \ra_{a,z}^\text{Fake} &= \frac{1}{2} + \alpha_z \left[ p \left(\la \wh{H}^- \ra_{a,z} - \frac{1}{2} \right) + (1-p) \left( \la \wh{H}^+ \ra_{a,z} - \frac{1}{2} \right) \right].
\end{align}
From these we see that Alice and Bob's $r$ parameter estimates are 
\begin{align}\label{Eq:rx+_fake}
\tilde{r}_x^+ &= \sqrt{2}\left[\la H^+\ra_{+,x}^\text{Fake} - \la H^-\ra_{+,x}^\text{Fake} \right]  
 = \alpha_x p_{basis}\left(2p_H - 1\right) \\ 
\tilde{r}_z^0 &= \sqrt{2}(\la H^+\ra_{0,z}^\text{Fake} + \la H^-\ra_{0,z}^\text{Fake} - 1) = \alpha_z p_{basis}.
\end{align}
Therefore the bit error rate estimated by Alice and Bob will be 
\begin{equation}\label{Eq:ErrorEstimate}
\tilde{\delta}_b = \frac{1}{2} - \frac{p_{basis}}{4}\left[\alpha_z + \alpha_x\left(2p_H-1\right)\right].
\end{equation}
From Eq. \eqref{Eq:rx+_fake}, we see that if Eve knows nothing about Bob's choice of weak measurement observable (i.e., $p_H = \frac{1}{2}$), then $\tilde{r}_x^+ = 0$ so that $\tilde{\delta}_X = \frac{1}{2}$, and hence $\tilde{\delta}_b$ will necessarily be at least $0.25$, which is well above the security bound. 

Let us suppose then that $p_H > \frac{1}{2}$. If we require that the protocol ensure that both $\tilde{\delta}_X$ and $\tilde{\delta}_Z$ are non-negative, then Eve is required to constrain her fake weak measurement results such that  
\begin{align}
\alpha_x &\leq \frac{1}{\left(2p_H - 1\right)p_{basis}} \\
\alpha_z &\leq \frac{1}{p_{basis}}.
\end{align}
This means that if Eve has some small amount of information about the weak measurement observable, i.e. $p_H > 1/2$, then she could reduce $\tilde{\delta}_b$ as low as desired even when she has no information about the source basis. However, unless Eve has near perfect information about the weak measurement observable, this comes at the cost of a drastic increase to the variance of the weak measurement results which can also be monitored independently by Alice and Bob. More explicitly, the variances of the fake measurement results are
\begin{align}
\label{Eq:FakeVarX-2}
\text{Var}[\Delta_\text{pointer}^\text{Fake}(H^{\pm},X)] &= \text{Var}[\alpha_x\Delta_{\pm}^\text{Eve}] = \left[ \frac{g_{Eve} \sigma_{Eve}}{\left(2p_H - 1\right)p_{basis}} \right]^2 \\
\label{Eq:FakeVarZ-2}
\text{Var}[\Delta_\text{pointer}^\text{Fake}(H^{\pm},Z)] &= \text{Var}[\alpha_z\Delta_{\pm}^\text{Eve}] = \left[ \frac{g_{Eve} \sigma_{Eve}}{p_{basis}} \right]^2. 
\end{align}

We now add to the protocol the requirement that Bob's weak measurement results have a variance that is consistent with a probability distribution that is specified for the weak measurement device. In terms of our model, this is a requirement on the WM black box that is under the control of Fred. Specifically, in our example, Bob will test the statistics of the weak measurement results obtained from Fred to verify they are consistent with a Gaussian distribution with variance equal to a specified value of $g^2 \sigma_{MD}^2$. This will require Eve to set
\begin{equation}
\text{Var}[\alpha_z\Delta_{\pm}^\text{Eve}] = \left[ \frac{g_{Eve} \sigma_{Eve}}{p_{basis}} \right]^2 = g^2 \sigma_{MD}^2.
\end{equation}
Since we have already noted that Eve will want to make $g_{Eve} \approx g$ in order to successfully fake the weak measurement results, this leaves Eve the flexibility to choose $\sigma_{Eve} = p_{basis}\sigma_{MD}$ to potentially hide her amplification factors. However, if Eve decreases $\sigma_{Eve}^2$ with a constant $g_{Eve}$, then Eve's weak measurements will become strong measurements as $g_{Eve}/\sigma_{Eve}$ approaches $1$. The effect of this is to add a component to the actual $\delta_Z$ error rate of the channel that Eve cannot remove. This will result in Alice and Bob underestimating the bit error rate which will cause the error reconciliation process to fail. Therefore, the ratio of Eve's coupling and weak measurement pointer must be less than or equal to the ratio of Alice and Bob's estimated coupling strength and expectation variance, 
\begin{equation}
\xi \equiv \left( \frac{g_{Eve}}{\sigma_{Eve}} \right)^2 \approx \left( \frac{g}{\sigma_{Eve}} \right)^2 \leq \left(\frac{g}{\sigma_{MD}}\right)^2, 
\end{equation}
with $\xi \ll 1$, in order to avoid introducing uncorrectable errors into the distributed raw key. In the case of white noise added to the weak measurement process and for asymptotically many signals, the approximation becomes equality and Eve is required to set $g_{Eve}=g$ and $\sigma_{Eve}=\sigma_{MD}$. Since the variances given in Equation \eqref{Eq:FakeVarX-2} and \eqref{Eq:FakeVarZ-2} must each equal $g^2 \sigma_{MD}^2$, Eve will only be able to fake the weak measurement results to satisfy the conditions for the protocol to certify the channel to the extent that she has some information about both the signal preparation basis and Bob's choice of $\wh{H}$ observable. 
 
To find the relationship between the bounds on Eve's information about the source basis and weak measurement observable and her control over the estimated bit error rate, note that the lower bound on the estimated bit error rate, $\tilde{\delta}_b^L$, is related to the upper bound of the estimated $r$ parameters by 
\begin{equation}
\tilde{\delta}_b^L = \frac{1}{2} - \frac{1}{4}\left( \tilde{r}_x^{+U} + \tilde{r}_z^{0U} \right).
\end{equation}
The upper bounds on the $r$ parameters are derived from the constraints on the variances of Eve's fake pointer results by
\begin{align}
\text{Var}[\alpha_x\Delta_{\pm}^\text{Eve}] &= \left[ \frac{\tilde{r}_x^{+U} g \sigma_{MD}}{\left(2p_H - 1\right)p_{basis}} \right]^2 = g^2 \sigma_{MD}^2 \\
\text{Var}[\alpha_z\Delta_{\pm}^\text{Eve}] &= \left[ \frac{\tilde{r}_z^{0U} g \sigma_{MD}}{p_{basis}} \right]^2 = g^2 \sigma_{MD}^2.
\end{align}
So that the $r$ upper bounds are 
\begin{align}
\tilde{r}_x^{+U} &= \left(2p_H - 1\right)p_{basis} \\
\tilde{r}_z^{0U} &= p_{basis}. 
\end{align}
From this we see that the lower bound on Alice and Bob's estimated bit error rate is 
\begin{equation}
\tilde{\delta}_b^L = \frac{1}{2} \left( 1- p_{basis}p_H \right).
\end{equation}
We therefore conclude that Attack Strategy 2 is only successful when $p_{basis}p_H > 1-2\delta_{sec}$, where $\delta_{sec}$ is the error security bound (e.g. $p_{basis}p_H > 0.78$ in the case of $\delta_{sec} = 11\%$), and therefore this strategy is strictly weaker than Strategy 3 (simple intercept-resend) since we have assumed that $p_H<1$. Additionally, when $p_H=1$, Attack Strategy 1 is strictly stronger than Attack Strategy 2 since it is independent of $p_{basis}$. 

If we allow the variance of the weak measurement pointer to have some imprecision, then in the weak measurement verification we can specify the ideal weak measurement pointer variance $\sigma_{MD}^2$ as well as an additional variance term $w^2$, e.g due to classical noise added to the pointer state.  In this more relaxed version of the variance test, Bob specifies both $\sigma_{MD}$ and $\sigma_{sec}$ for the weak measurement implementation, with $\sigma_{sec}^2=\sigma_{MD}^2+w^2$. In this more realistic case, the equations for the $r$ upper bounds become
\begin{align}
\text{Var}[\alpha_x\Delta_{\pm}^\text{Eve}] &= \left[ \frac{\tilde{r}_x^{+U} g \sigma_{MD}}{\left(2p_H - 1\right)p_{basis}} \right]^2 = g^2 \sigma_{sec}^2\\
\text{Var}[\alpha_z\Delta_{\pm}^\text{Eve}] &= \left[ \frac{\tilde{r}_z^{0U} g \sigma_{MD}}{p_{basis}} \right]^2 = g^2 \sigma_{sec}^2.
\end{align}

From these we arrive at a modified lower bound on the estimated bit error rate for the non-ideal weak measurement implementation case
\begin{equation}
\tilde{\delta}_b^L = \frac{1}{2} \left[1 -  \left(\frac{\sigma_{sec}}{\sigma_{MD}}\right) p_{basis}p_H \right].
\end{equation}
If $p_{basis}$ and $p_H$ are known (or at least upper bounds of these parameters can be estimated with high confidence), then security of the protocol is enforced by choosing $\sigma_{sec}$ such that $\tilde{\delta}_b^L$ is necessarily larger than the QBER security threshold $\delta_{sec}$. It is rather striking to see just how much weak measurement imprecision is permitted when $p_{basis}$ and $p_H$ are close to $1/2$, while still maintaining security. For example, if $p_{basis}p_H = 0.35$ and we use the Shor-Preskill \cite{ShorPreskill} QBER bound of $\delta_{sec}=0.11$, then $\sigma_{sec}$ can be over twice $\sigma_{MD}$ before $\delta_{sec}$ exceeds $\tilde{\delta}_b^L$.

\section{The Full Protocol\label{App:Protocol}}
In Table \ref{Tab:Protocol} we provided a sketch of a QKD protocol that performs parameter estimation with weak measurements. In this appendix, we provide a more detailed version that incorporates all of the subtleties that were discussed in the article. The primary difference between the WM-QKD protocol and most other QKD protocols is the parameter estimation, so we describe that subroutine independently, after writing out the protocol. We will implicitly assume that decoy states have been implemented as in Section \ref{Subsec:ImperfectSources} in order to estimate the appropriate parameters.

\subsection{The Protocol}
\begin{itemize}
\item[1)]	Alice creates two uniformly random binary strings, $s_A$ and $b$, of length $N$, corresponding to the bit and basis, respectively. If $b_i = 0$, then Alice encodes her bit ${s_A}_i$ in the $Z$ basis; otherwise she encodes it in the $X$ basis. She then transmits each signal to Bob.

\item[2)]	Bob creates a uniformly random length $N$ binary string $h$ which he uses to choose his weak measurement observable on each signal. If $h_i = 0$, he weakly measures $\wh{H}^+$ on the $i^{th}$ signal; otherwise he weakly measures $\wh{H}^-$. For each signal received, Bob weakly measures the signal according to $h$, then strongly measures the signal in the $Z$ basis. Bob records his weak measurement result in a (real-valued) string $\omega$ and his strong measurement result in a string $s_B$ that takes the values $0$, $1$, and $\bot$, corresponding to measurement outcome $+1$, $-1$, and no signal received, respectively.

\item[3)]	After the $N^{th}$ signal, Bob announces which indices in $s_B$ have the value $\bot$. Alice and Bob remove these indices from their respective strings $s_A$, $b$, $h$, $\omega$, and $s_B$.

\item[4)]	Bob publicly announces the strings $h$ and $\omega$. Alice uses these strings to perform the parameter estimation subroutine.

\item[5)] Alice either announces the abort signal, or the QBER, along with her string $b$.

\item[6)]	If the protocol is not aborted, then Alice and Bob both remove the indices $i$ from their respective strings, $s_A$ and $s_B$, for which $b_i = 1$. Their reduced strings now correspond to their sifted keys.

\item[7)]	Alice and Bob perform classical error reconciliation and privacy amplification as usual, to obtain a secure shared key from their sifted keys.

\end{itemize}

\subsection{Parameter Estimation Subroutine}
We assume Alice has the following threshold parameters: $\delta_{sec}$, $g_{sec}$, and $\sigma_{sec}^2$. We additionally assume she is able to obtain the fraction $d(\mu)$ of signals due to dark counts from the decoy states, as in Eq. \eqref{Eq:DarkCountFraction}.

\begin{itemize}
\item[(1)]	For each length-3 binary string $\ov{a} \in \mbb{Z}_2^3$ Alice creates a substring $\mu_{\ov{a}}$ of the weak measurement string $\omega$ received from Bob according to
\begin{equation}
\omega_i \in \mu_{({s_A}_i, b_i, h_i)}.
\end{equation}
This conditions the weak measurement results of each observable at each initial state.

\item[(2)]	Alice calculates the averages $\ov{\mu_{\ov{a}}}$ of each of these strings, as well as the variance $\sigma_{\ov{a}}^2$ of each.

\item[(3)]	Using Eqs. \eqref{Eq:g+} and \eqref{Eq:g-}, and the estimate $d(\mu)$ of the signal dark count fraction, Alice calculates $g^+$ and $g^-$.

\item[(4)] Alice then calculates all 8 expectation values, from which she obtains her estimates of $\delta_X$ and $\delta_Z$.
 
\item[(5)]	Alice certifies the weak measurements by verifying that 
\begin{align}
\delta_X, \ \delta_Z &\geq 0,\\
g^+, \ g^- &\leq g_{sec}, \\
\sigma_{\ov{a}}^2 &\leq \sigma_{sec}^2, \text{ for each } \ov{a}, \text{ and}\\
\sigma_{\ov{a}}^2 &\approx \sigma_{\ov{a}'}^2, \text{ for all } \ov{a}, \ov{a}'.
\end{align}

\item[(6)]	If any of the above relationships fail, Alice sends an abort signal. Otherwise, she calculates the QBER according to
\begin{equation}
\textup{QBER} = \delta_b + (1-d(\mu))\delta_{wm},
\end{equation}
where $\delta_b$ is the channel bit error rate, calculated from, e.g., Eq. \eqref{Eq:QBER_darkcounts}, $\delta_{wm}$ is the weak measurement induced error rate calculated in Section \ref{Sec:WM-disturbance}, and $d(\mu)$ is the fraction of signals due to dark counts, given by Eq. \eqref{Eq:DarkCountFraction}. Alice then performs the security check by verifying
\begin{equation}
\textup{QBER} \leq \delta_{sec}.
\end{equation}
If the above equation fails, Alice returns an abort symbol; otherwise she returns the QBER.
\end{itemize}

\end{document}